\documentclass{PoS}
\usepackage{epsfig,amsmath,cite}

\def\iscol#1#2{{\buildrel#1\parallel#2\over\longrightarrow}}
\def\beq{\begin{equation}}
\def\eeq{\end{equation}}
\def\beqa{\begin{eqnarray}}
\def\eeqa{\end{eqnarray}}

\title{Long-distance singularities
 in multi-leg scattering amplitudes\thanks{Preprint numbers: Edinburgh 2016/10, CERN-TH-2016-141, CP3-16-31}}

\ShortTitle{IR singularities at three-loops}

\author{\speaker{Einan Gardi}\\
        Higgs Centre for Theoretical Physics, School of Physics and Astronomy, 
The University of Edinburgh, Edinburgh EH9 3FD, Scotland, UK\\
        E-mail: \email{Einan.Gardi@ed.ac.uk}}

\author{\O{}yvind Almelid\\
        Higgs Centre for Theoretical Physics, School of Physics and Astronomy, 
The University of Edinburgh, Edinburgh EH9 3FD, Scotland, UK\\
        E-mail: \email{Oyvind.Almelid@ed.ac.uk}}

\author{Claude Duhr\\
{CERN Theory Division, 1211 Geneva 23, Switzerland}\\
{Center for Cosmology, Particle Physics and Phenomenology (CP3),
Universit\'{e} Catholique de Louvain, 1348 Louvain-La-Neuve, Belgium\thanks{On leave from the ``Fonds National de la Recherche Scientifique'' (FNRS), Belgium.}} \\
E-mail:  \email{Claude.Duhr@cern.ch}}

\abstract{We report on the recent completion of the three-loop calculation of the soft anomalous dimension in massless gauge-theory scattering amplitudes. This brings the state-of-the-art knowledge of long-distance singularities in multi-leg QCD amplitudes with any number of massless particles to three loops. The result displays some novel features: this is the first time non-dipole corrections appear, which directly correlate the colour and kinematic degrees of freedom of four coloured partons. We find that non-dipole corrections appear at three loops also for three coloured partons, but these are independent of the kinematics. The final result is remarkably simple when expressed in terms of single-valued harmonic polylogarithms, and it satisfies several non-trivial constraints. In particular, it is consistent with the high-energy limit behaviour and it satisfies the expected factorization properties in two-particle collinear limits.      }

\FullConference{Loops and Legs in Quantum Field Theory\\
		24-29 April 2016\\
		Leipzig, Germany}

\begin{document}

\section{Introduction}

Long-distance singularities are a central feature of gauge-theory scattering amplitudes, and a 
detailed understanding of their structure is key to precision collider physics~\cite{Gatheral:1983cz,Frenkel:1984pz,Korchemsky:1985xj,Magnea:1990zb,Korchemsky:1993uz,Korchemsky:1993hr,Catani:1996vz,Sterman:1995fz,Catasterm,Kidonakis:1998nf,Bonciani:2003nt,Dokshitzer:2005ig,Aybat:2006mz,Dixon:2008gr,Dixon:2009gx,Gardi:2009qi,Becher:2009cu,Gardi:2009zv,Dixon:2009ur,Bret:2011xm,Caron-Huot:2013fea,Erdogan:2014gha,Ahrens:2012qz,Gehrmann:2010ue,Naculich:2013xa,Kidonakis:2009ev,Mitov:2009sv,Becher:2009kw,Beneke:2009rj,Czakon:2009zw,Ferroglia:2009ep,Ferroglia:2009ii,Chiu:2009mg,Mitov:2010xw,Gardi:2013saa,Henn:2013tua,Gardi:2010rn,Mitov:2010rp,Gardi:2011wa,Gardi:2011yz,Gardi:2013ita}. Owing to their factorization properties, the singularities are largely independent of the hard scattering process. Furthermore, they exponentiate and can therefore be compactly summarised by the so called soft anomalous dimension.

Until recently, the soft anomalous dimension for the scattering of any number of massless coloured particles was known to two loops. To this order it admits a remarkably simple structure consisting of a sum over colour dipoles formed by any pair of external legs~\cite{Catasterm,Aybat:2006mz,Gardi:2009qi,Becher:2009cu,Gardi:2009zv}. 
In this talk we report on the recent computation of the three-loop corrections to the soft anomalous dimension~\cite{Almelid:2015jia}. The calculation we performed confirmed the expectation~\cite{Gardi:2009qi,Becher:2009cu,Gardi:2009zv,Dixon:2009ur,Bret:2011xm,Caron-Huot:2013fea,Ahrens:2012qz} that three-loop corrections depart from the above dipole structure, and correlate between the kinematic and colour degrees of freedom of up to four partons. We find that a non-vanishing correction appears already for three coloured partons, but it is a constant, involving no kinematic dependence. The new three-loop result also contributes to understanding factorization properties of scattering amplitudes in the collinear and high-energy limits. 

\section{Factorization at fixed-angles and the soft anomalous dimension}

We are interested in the infrared (IR) structure of a scattering amplitude for $n$ massless partons. Given external legs with momenta $p_i$, for $i=1..n$, where $p_i^2=0$, 
we consider the kinematic limit of \emph{fixed-angle scattering}, where all Lorentz invariants  $p_i\cdot p_j$ are taken large. 
Infrared  singularities (both soft and collinear) can then be factorized as follows 
\begin{equation}
\label{amp_factorization}
{\cal M}_n\left(\left\{p_i\right\}, \alpha_s \right) =
Z_n \left(\left\{p_i\right\}, \mu, \alpha_s \right) 
{\cal H}_n \left(\left\{p_i\right\}, \mu, \alpha_s \right)\,, 
\end{equation}
where $\mu$ is a factorization scale, $\alpha_s\equiv \alpha_s(\mu^2)$ is the renormalised $D$-dimensional running coupling, ${\cal H}_n$ is a {finite} hard scattering function, and $Z_n$ is an operator in colour space collecting all IR singularities as poles in the dimensional regularization parameter $\epsilon = (4-D)/2$. These singularities originate in loop momenta becoming either soft or collinear to any of the scattered partons (see e.g. Ref.~\cite{Sterman:1995fz}). Collinear singularities depend on the spin and momentum of that particle, and decouple from the rest of the process; their contribution is known to three-loops~\cite{Moch:2005tm,Gehrmann:2010ue}, and will not be discussed here in detail.  
In contrast, soft singularities are independent of the spin, but they depend on the relative directions of motion and the colour degrees of freedom of all scattered particles. Hence, soft singularities are sensitive to the colour flow in the entire process. Nevertheless, they are significantly simpler than finite contributions to the amplitude, opening a unique possibility to explore multi-leg gauge-theory amplitudes at the multi-loop level.

The simplification of the soft limit is apparent already at the level of the Feymann rules: emission of a soft gluon with momentum $k$ off an energetic particle with momentum $p_i\gg k$, taken at leading order in the soft gluon momentum, amounts to a factor of $g_s {\bf T}_i^a \frac{p_i^{\mu}}{p_i\cdot k +{\rm i}0} = g_s {\bf T}_i^a \frac{\beta_i^{\mu}}{\beta_i\cdot k +{\rm i}0}$, where we replaced the momentum of the emitting particle by its four-velocity, emphasising the \emph{rescaling symmetry} of this Feynman rule. This symmetry is responsible for the main features of soft singularities. 
The soft approximation can be equivalently formulated in configuration space, as emission from a Wilson line following the 
classical trajectory of the particle with momentum $p_i$ and carrying the same colour charge:
\begin{equation}
\Phi_{\beta_i}\,\equiv\,
{\cal P}\exp\left[{\mathrm i}g_s \int_0^{\infty}dt \beta_i\cdot {A}^a(t\beta_i) {\bf T}_i^a\right]\,,
\end{equation}
where ${\cal P}$ orders the colour matrices along the path. 
To avoid collinear singularities we perform our calculation with non-lightlike velocities $\beta_i^2\neq 0$. Considering fixed-angle scattering of $n$ legs, soft singularities are fully captured by the following Wilson-line correlator, the so-called \emph{soft function},
\begin{equation}
\label{Soft}
S\left(\left\{\gamma_{ij}\right\};\mu\right)\equiv \left< {\rm T}\left(\Phi_{\beta_1}\otimes \Phi_{\beta_2}\ldots \otimes \Phi_{\beta_i}\otimes \ldots \Phi_{\beta_n}\right)\right>
\end{equation}
where the kinematic dependence appears through cusp angles, $\gamma_{ij}\equiv 2\beta_i\cdot\beta_j/\sqrt{\beta_i^2\beta_j^2}$, which are invariant under velocity rescaling.

The factor $Z_n$ containing all soft and collinear singularities in Eq.~(\ref{amp_factorization}) can be written as a solution of a renormalization-group equation as
\begin{equation}\begin{split}
\label{sumodipoles}
\! Z_n&   = {\cal P}
  \exp \Bigg\{\! -\frac12 \int_0^{\mu^2}\! \frac{d \lambda^2}{\lambda^2} \Gamma_n\left(\left\{p_i\right\}, \lambda,\alpha_s(\lambda^2)\right)\! \Bigg\}\,,
\end{split}\end{equation}
where $\Gamma_n$ is the so-called \emph{soft anomalous dimension matrix} for multi-leg scattering, and ${\cal P}$ stands for path-ordering of the matrices according to the order of scales~$\lambda$.  $\Gamma_n$ itself is finite, and IR singularities are generated in Eq.~\eqref{sumodipoles} 
through the dependence of $\Gamma_n$ on the $D$-dimensional coupling,
which is integrated over the scale down to zero momentum. Factorization and the rescaling symmetry of the Wilson line velocities~\cite{Gardi:2009qi,Becher:2009cu,Gardi:2009zv} put stringent constraints on the functional form of $\Gamma_n$, which through three loops, must take the form
\begin{align}
\label{Gamma}
\begin{split}
\Gamma_n
\left(\left\{p_i\right\}, \lambda\right) 
=
\Gamma_{n}^{\rm dip.}
\left(\left\{p_i\right\}, \lambda\right) 
+\Delta_n\left(\left\{\rho_{ijkl}\right\}\right)\,,
\end{split}
\end{align}
with
\beqa
\label{dipole_formula}
\Gamma_{n}^{\rm dip.} \left(\left\{p_i\right\}, \lambda\right) = & -&\frac{1}{2} \widehat{\gamma}_K \left( \alpha_s \right) 
  \sum_{i<j}  \log \left(\frac{-s_{ij}}{\lambda^2} \right)
  {\bf T}_i \cdot {\bf T}_j  + \sum_{i = 1}^n
  \gamma_{J_i} \left(\alpha_s\right) \,,
\eeqa
where $- s_{ij} = 2 \left\vert p_i \cdot p_j \right \vert e^{ - {\rm i}
\pi \lambda_{ij}}$, with $\lambda_{i j} = 1$ if partons
 $i$ and $j$ 
both belong to either the initial or the final state and $\lambda_{i j} = 0$ 
otherwise; ${\bf T}_i$ is the colour generator in the representation of 
parton~$i$, acting on the colour indices of the amplitude as described 
in Ref.~\cite{Catani:1996vz}; $\widehat{\gamma}_K (\alpha_s)$ is the universal cusp anomalous dimension~\cite{Korchemsky:1985xj,Grozin:2014hna,Moch:2004pa}, with the quadratic Casimir of the appropriate representation scaled out\footnote{Casimir scaling of the cusp anomalous dimension holds through three loops~\cite{Moch:2004pa}; it may be broken by 
quartic Casimirs starting at four loops.}; $\gamma_{J_i}$ 
are the anomalous dimensions of the fields associated with external 
particles, which govern hard collinear singularities, currently known to three loops~\cite{Moch:2005tm,Gehrmann:2010ue}. Equation~\eqref{dipole_formula} is known as the \emph{dipole formula}, and captures the entirety of the soft anomalous dimension up to two loops. 
Finally, $\Delta_n\left(\left\{\rho_{ijkl}\right\}\right)$ represents the correction going beyond the dipole formula, which starts at three loops,
\begin{equation}
\Delta_n\left(\left\{\rho_{ijkl}\right\}\right) = \sum_{\ell=3}^\infty\left(\frac{\alpha_s}{4\pi}\right)^\ell \Delta_n^{(\ell)}\left(\left\{\rho_{ijkl}\right\}\right)\,.
\end{equation}
and depends on the kinematics via \emph{conformally-invariant cross ratios} (CICRs), 
\beq\label{eq:CICR}
\rho_{ijkl}\equiv\frac{(-s_{ij})(-s_{kl})}{(-s_{ik})(-s_{jl})}\,
=\, \frac{\gamma_{ij}\,\gamma_{kl}}{\gamma_{ik}\, \gamma_{jl}}
\,,
\eeq
which are invariant under a rescaling of any of the momenta. In the following we report on the calculation of the three-loop function $\Delta_n^{(3)}\left(\left\{\rho_{ijkl}\right\}\right)$.

With the exception of hard collinear singularities ($\gamma_{J_i}(\alpha_s)$ in Eq.~(\ref{dipole_formula})), one may compute the soft anomalous dimension $\Gamma_n\left(\left\{p_i\right\}, \lambda\right)$ to any order through the renormalization of the soft function in Eq.~(\ref{Soft}): in dimensional regularization, loop corrections to the soft function are scale-less integrals, which vanish in the absence of a cutoff. Hence, one may directly infer the infrared poles in $\epsilon$ from the ultraviolet ones. This calculation strategy has marked advantages over the alternative of extracting the infrared poles from an amplitude, since one never needs to evaluate finite corrections, and one may make direct use of the known iterative structure of renormalization along with the exponentiation properties of Wilson line correlators~\cite{Gatheral:1983cz,Frenkel:1984pz,Gardi:2010rn,Mitov:2010rp,Gardi:2011wa,Gardi:2011yz,Gardi:2013ita,Gardi:2013saa}.

\begin{figure*}[!t]
\begin{center}
\scalebox{.25}{\includegraphics{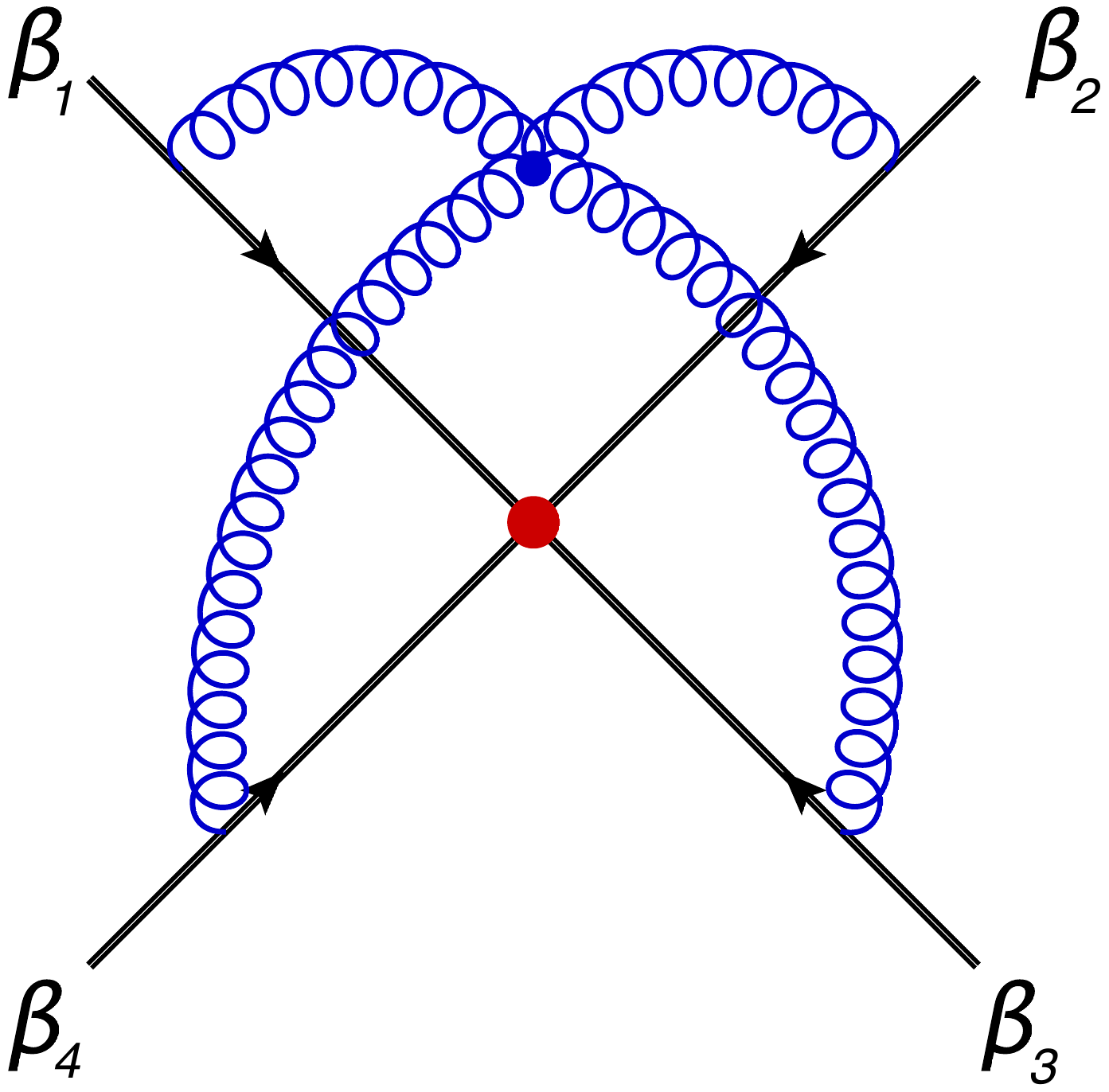}}\hspace*{10pt}
\scalebox{.25}{\includegraphics{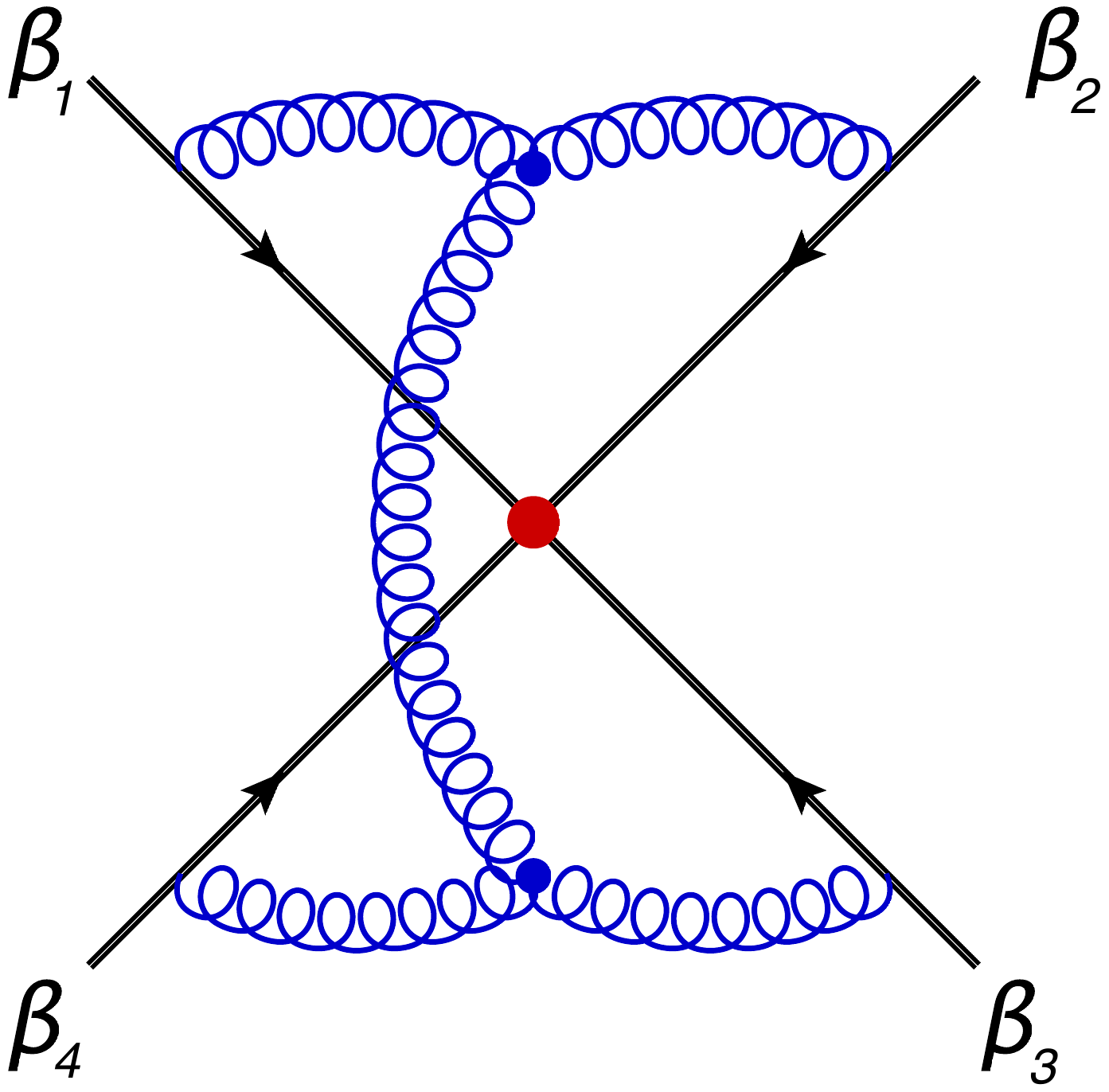}}\hspace*{10pt}
\scalebox{.25}{\includegraphics{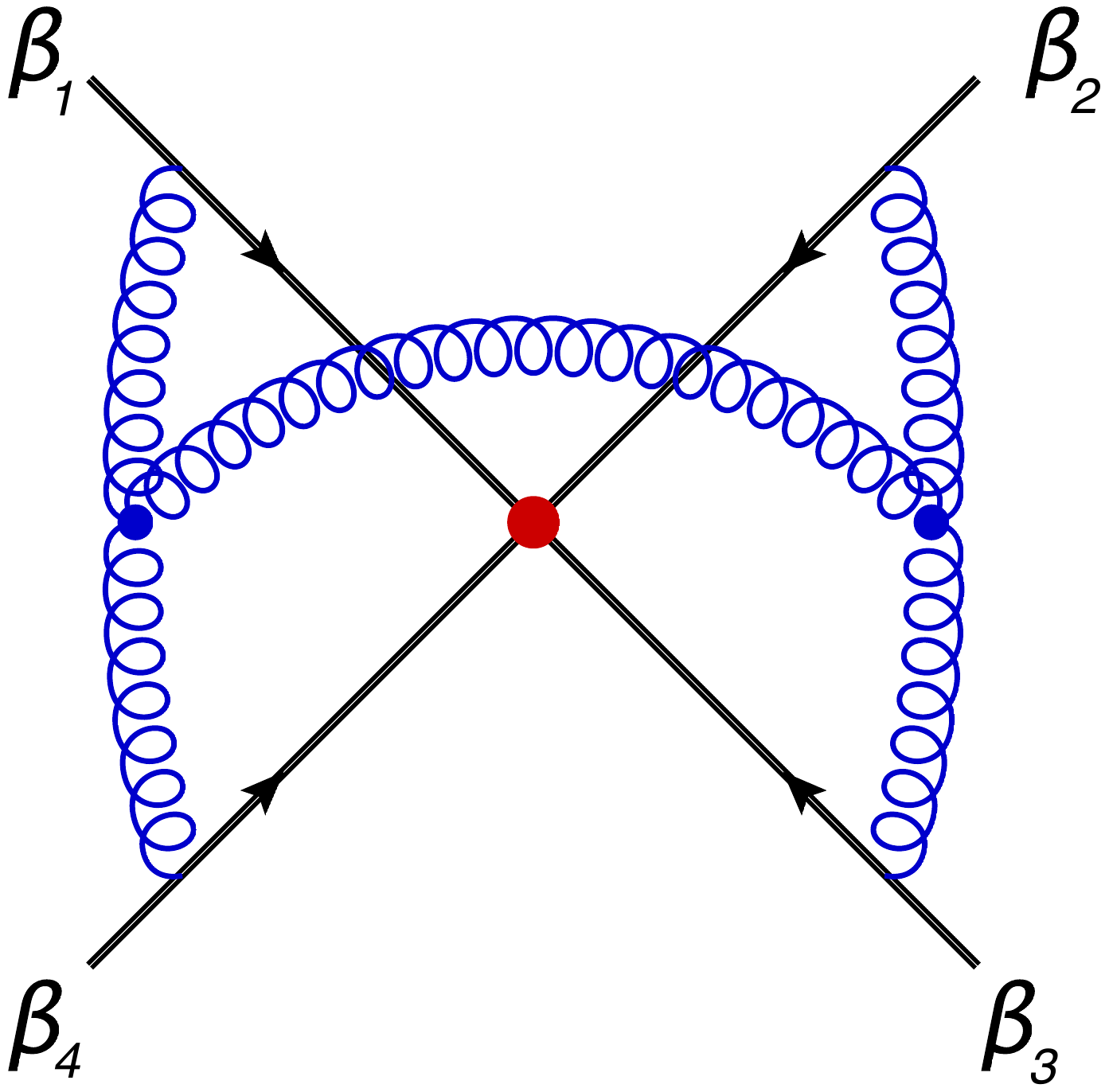}}\hspace*{10pt}
\scalebox{.25}{\includegraphics{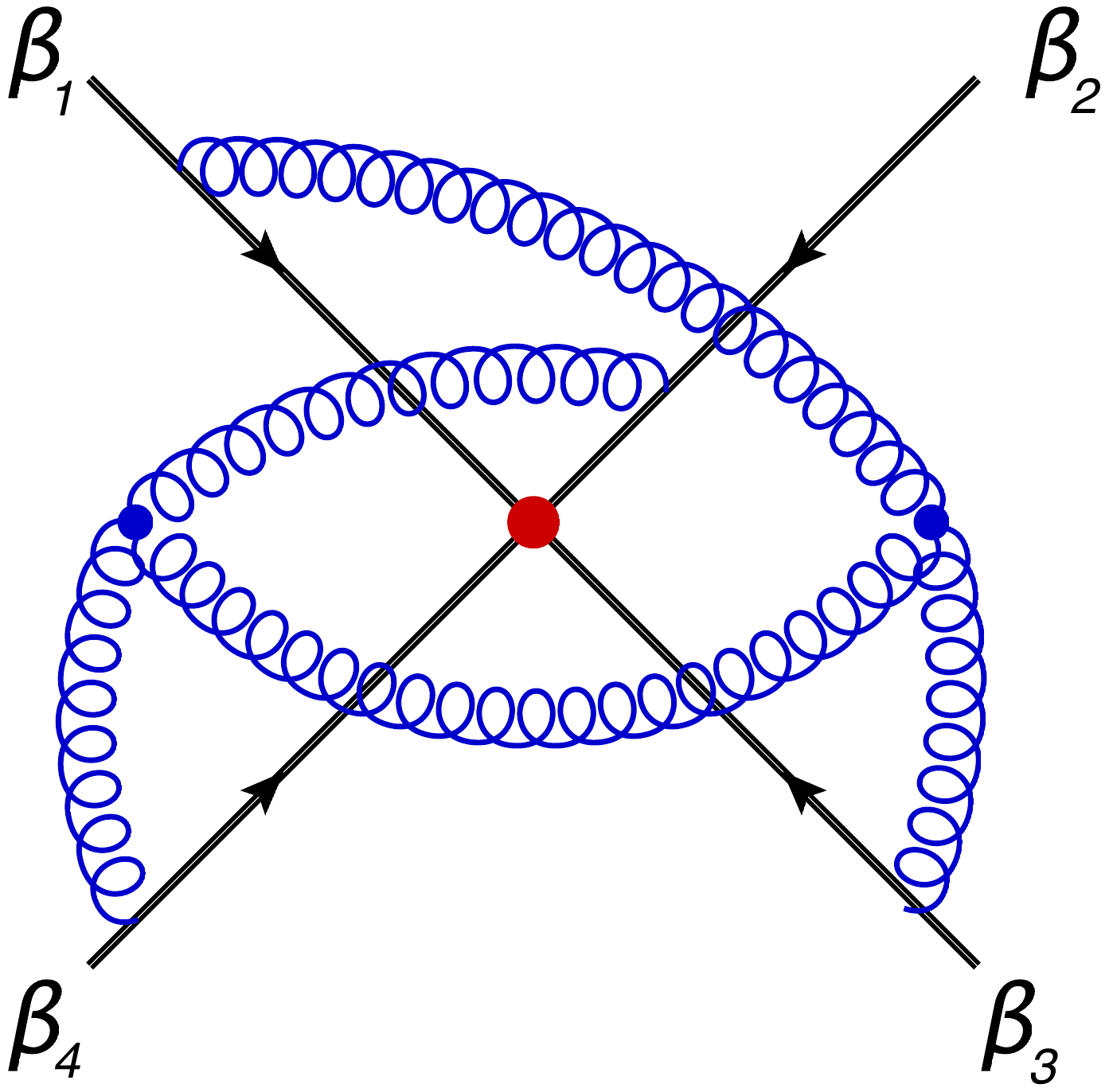}}\hspace*{10pt}
\caption{All connected 3-loop webs connecting four Wilson lines.}
\label{4lines_connected}
\end{center}
\end{figure*}

\begin{figure*}[t]
\begin{center}
\scalebox{.23}{\includegraphics{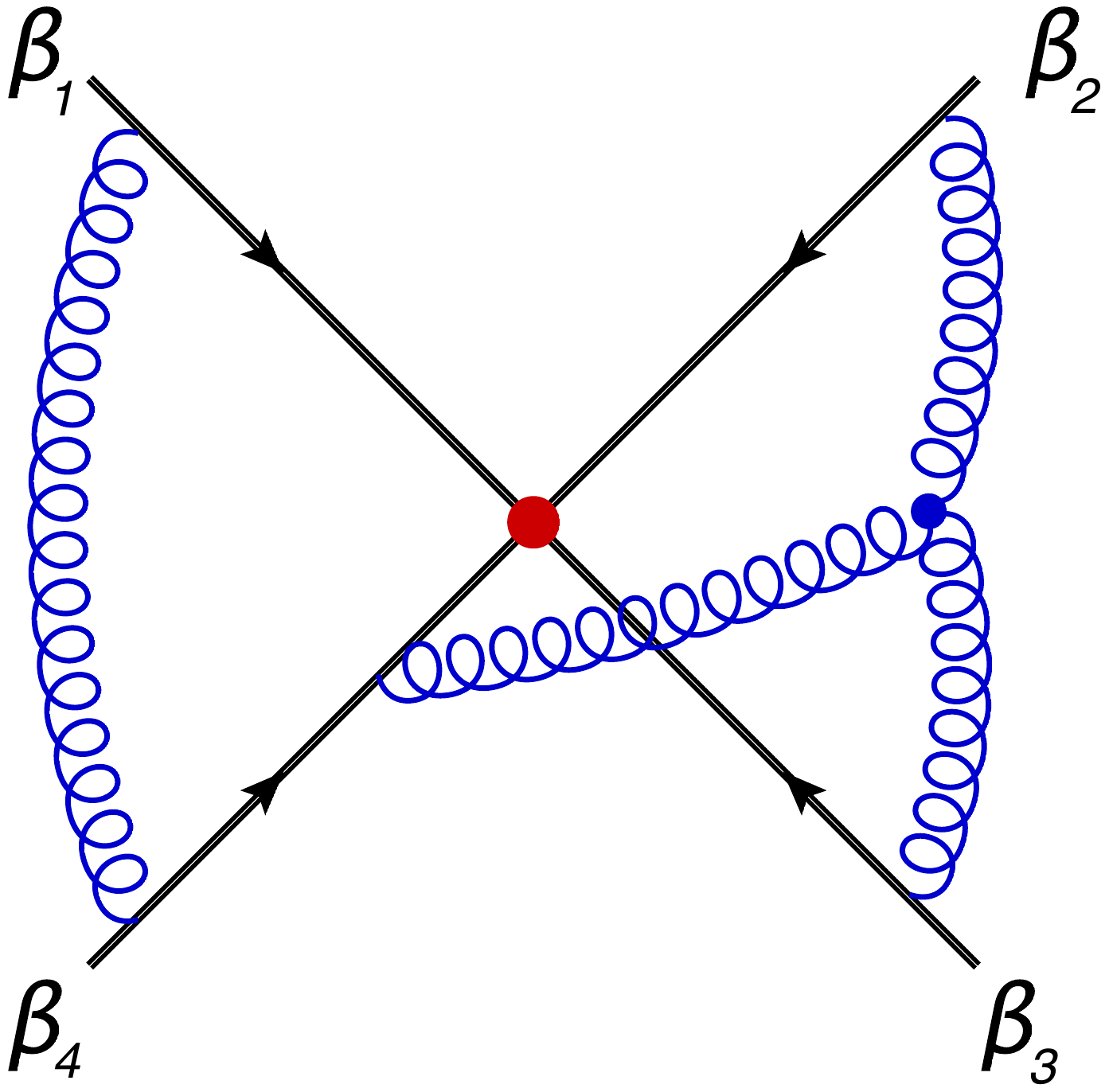}}
\hskip 1.5cm
\scalebox{.23}{\includegraphics{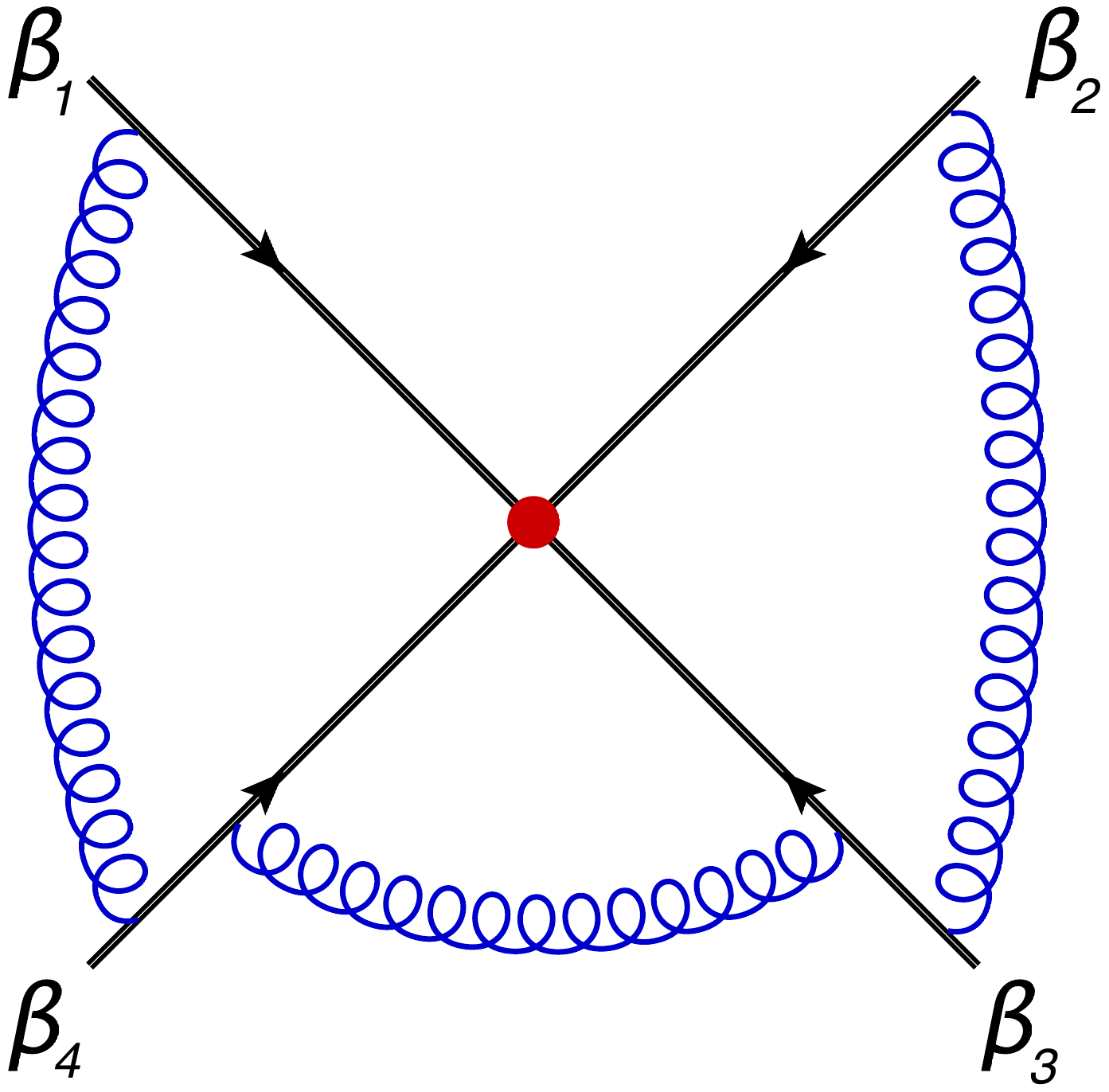}}
\hskip 1.5cm
\scalebox{.23}{\includegraphics{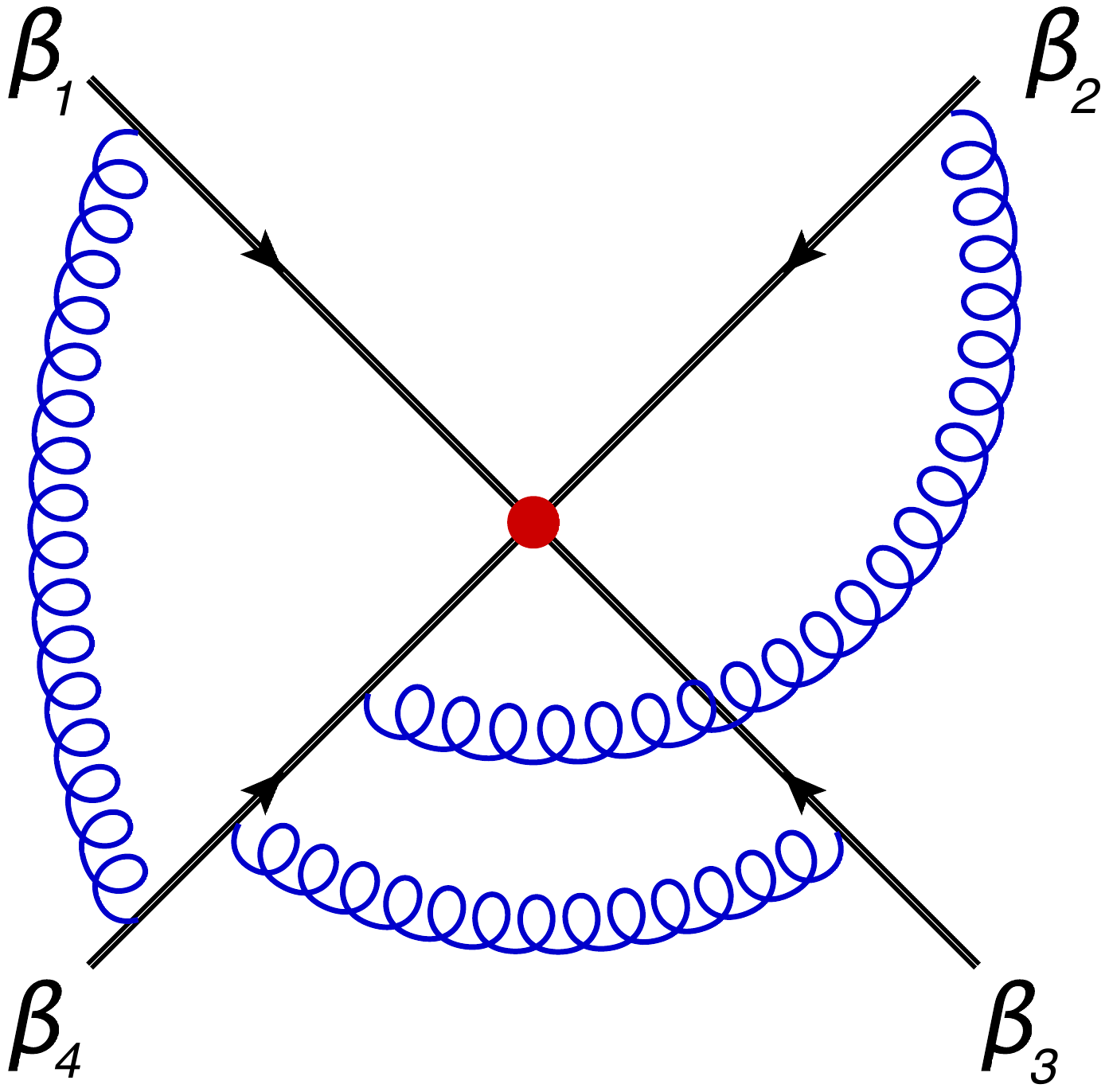}}
\caption{Representative non-connected 3-loop diagrams of webs connecting four Wilson lines.}
\label{4lines_non-connected}
\end{center}
\end{figure*}

\begin{figure*}[!t]
\begin{center}
\scalebox{.25}{\includegraphics{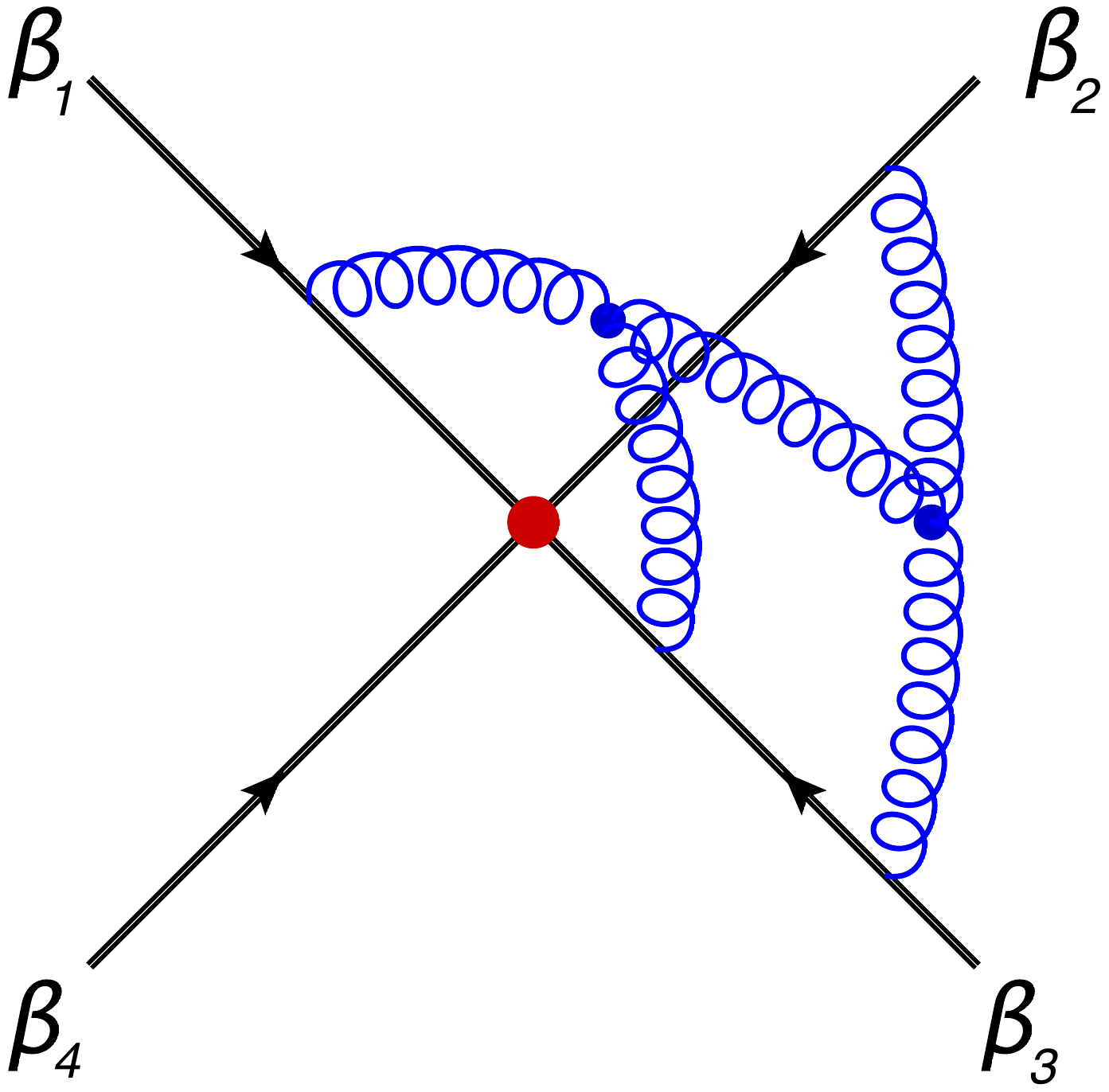}}\hspace*{10pt}
\scalebox{.25}{\includegraphics{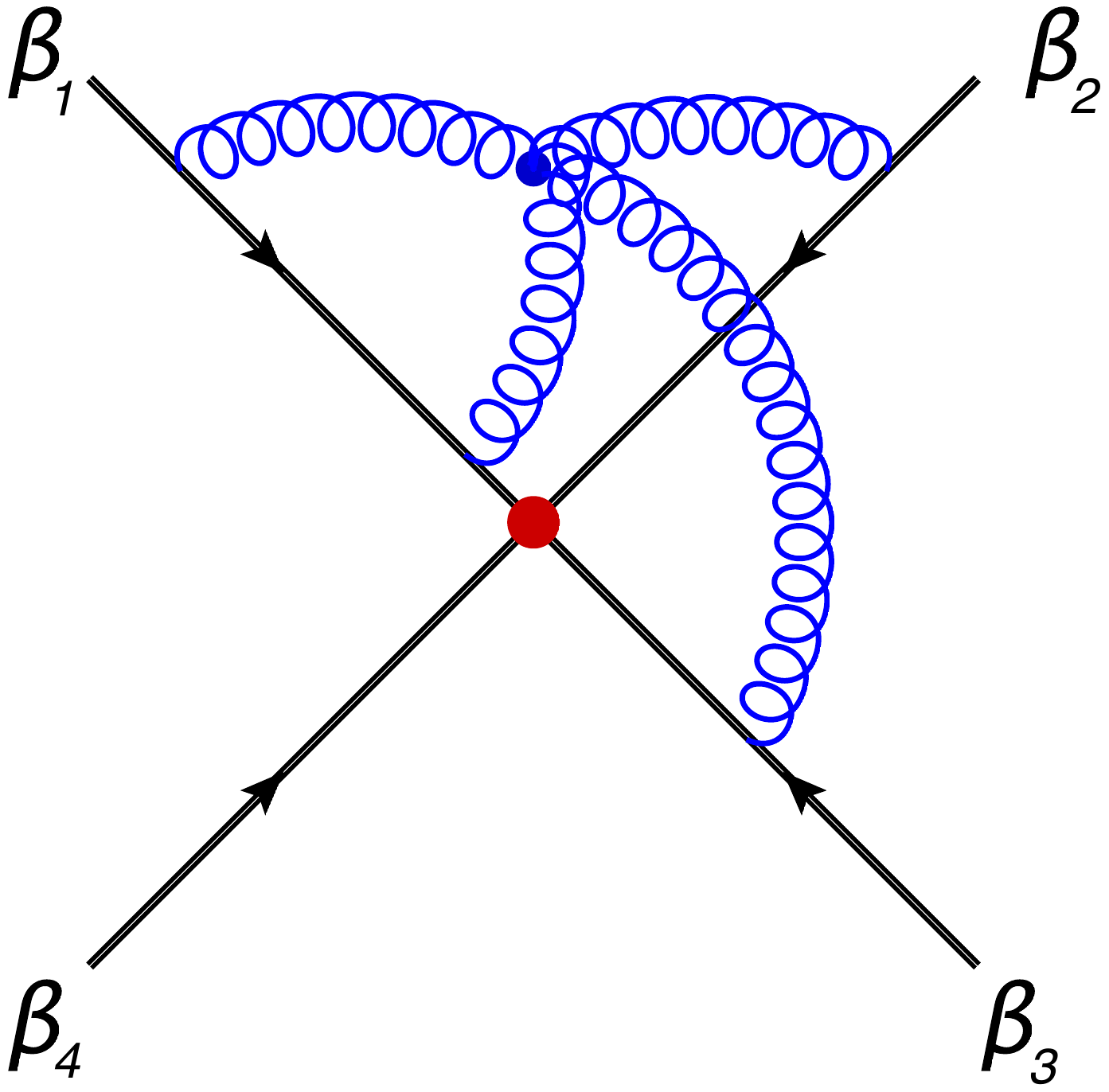}}\hspace*{10pt}
\scalebox{.25}{\includegraphics{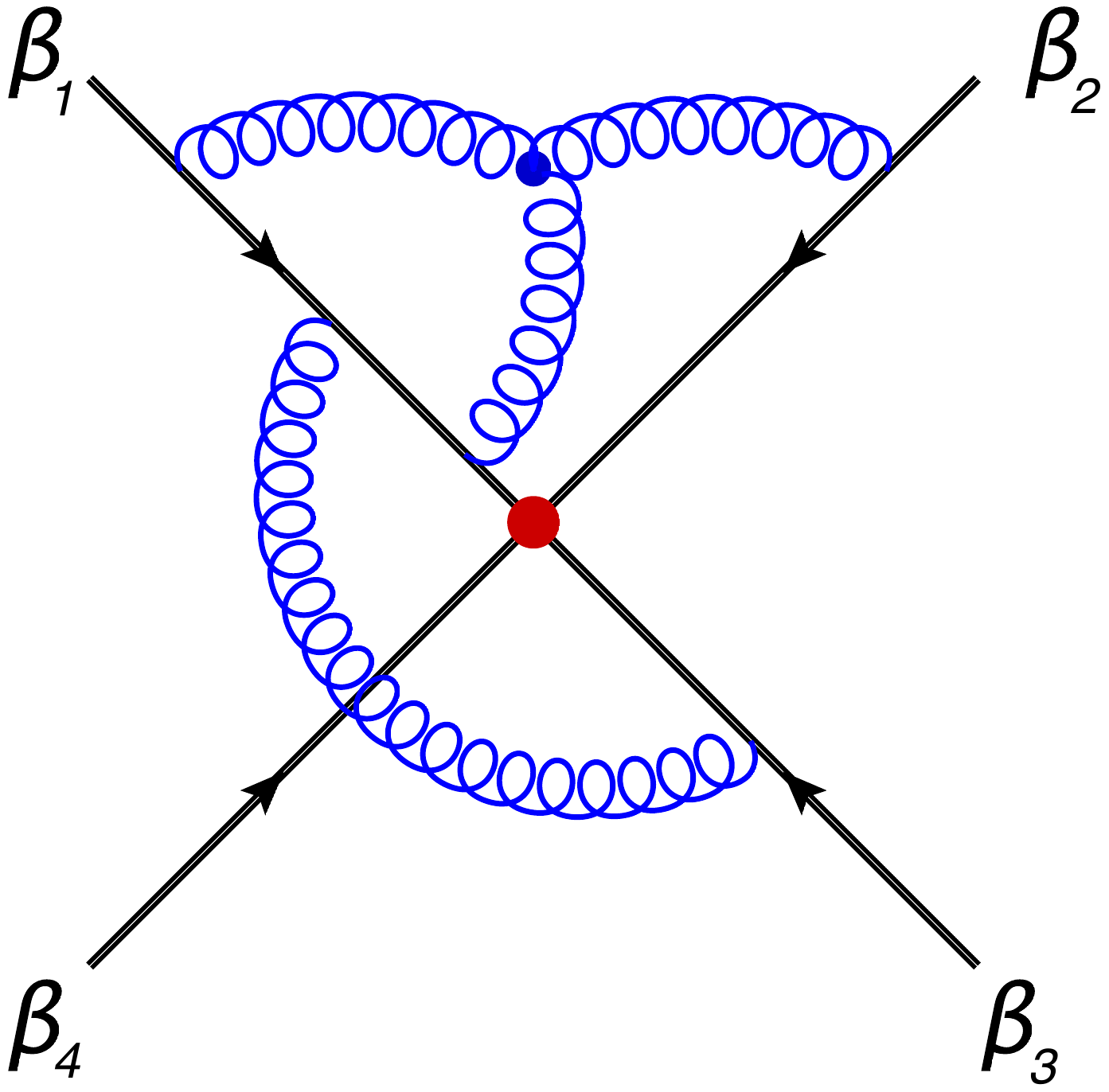}}\hspace*{10pt}
\scalebox{.25}{\includegraphics{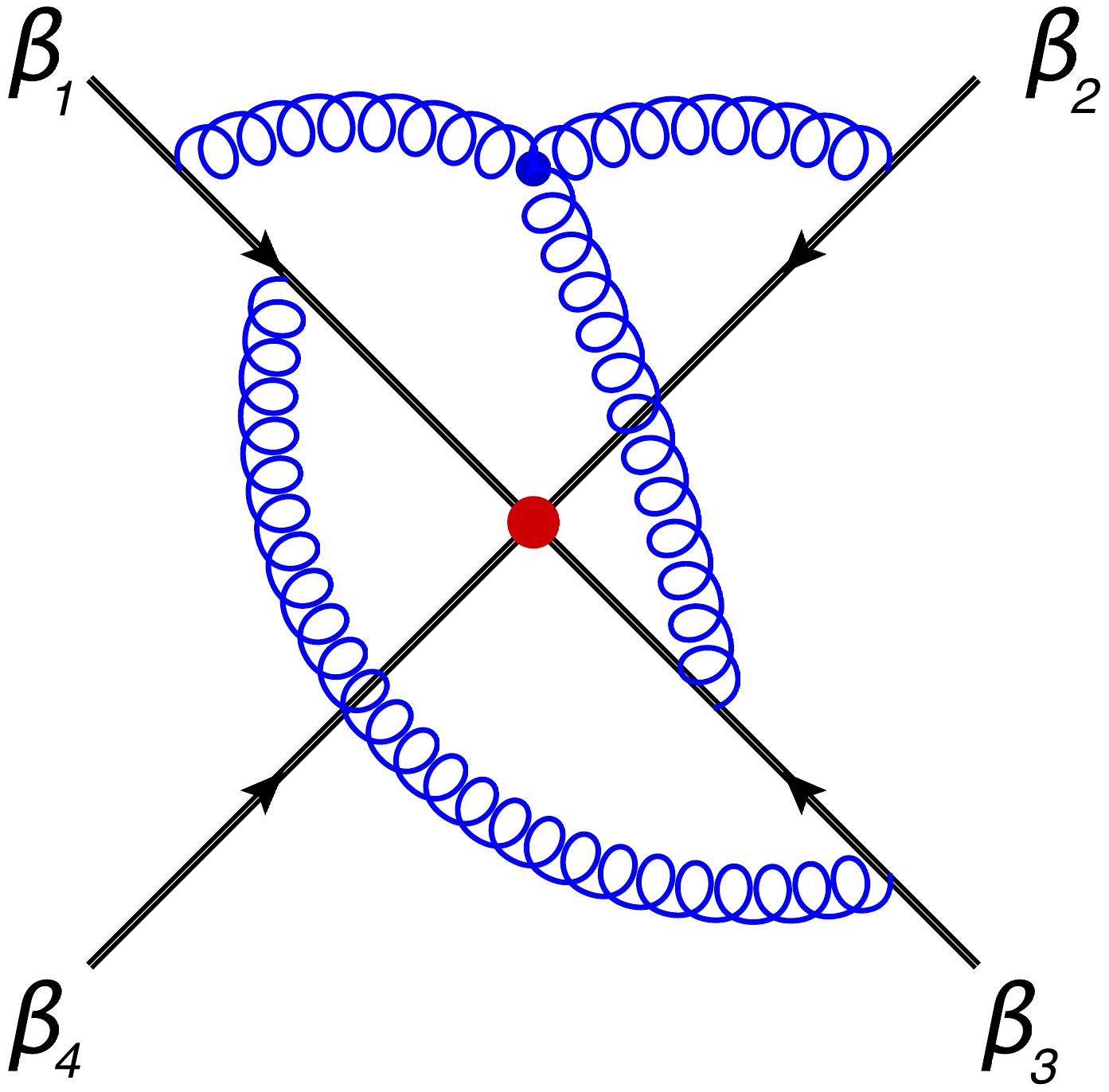}}\hspace*{10pt}\\
\scalebox{.25}{\includegraphics{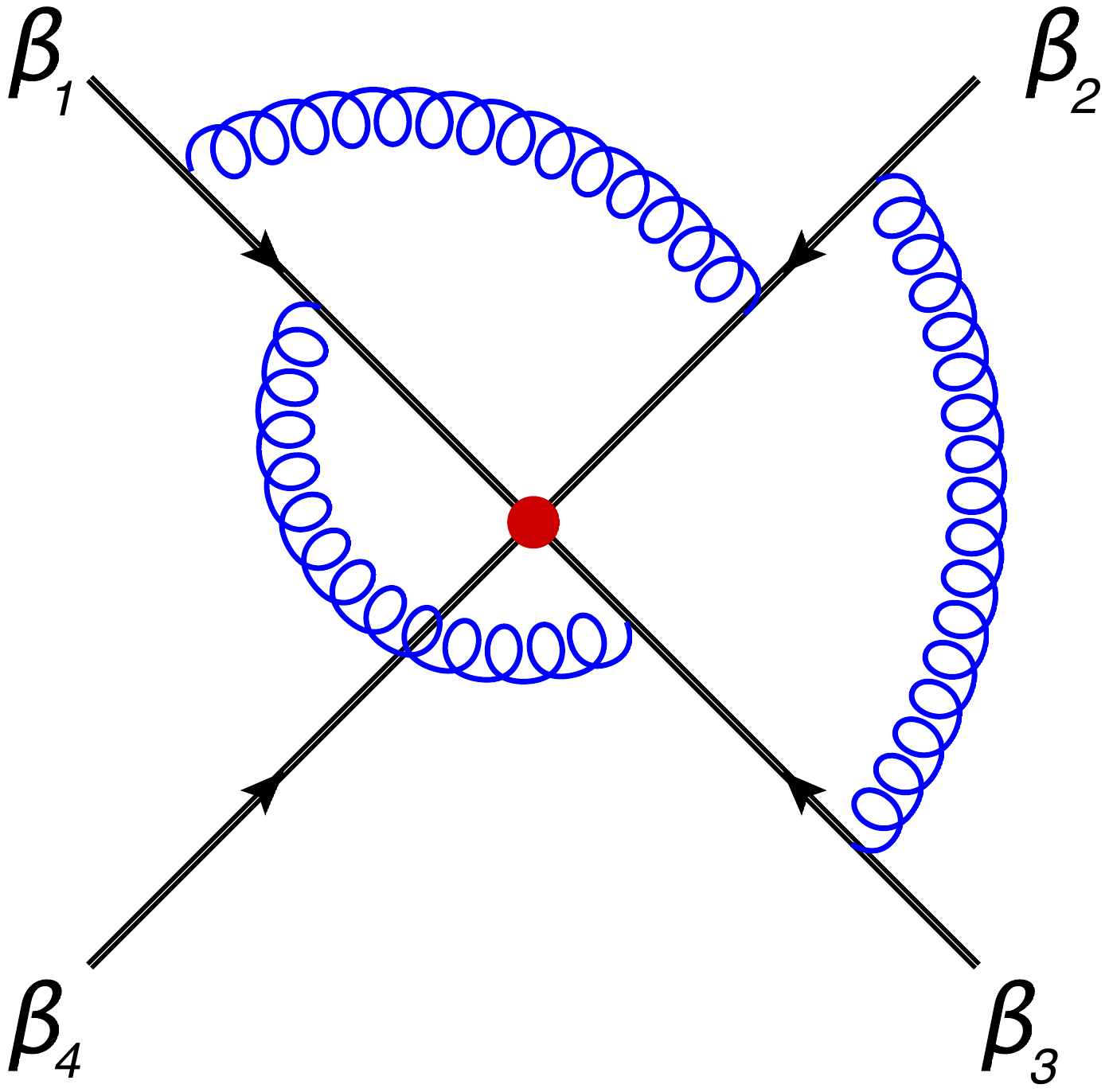}}\hspace*{10pt}
\scalebox{.25}{\includegraphics{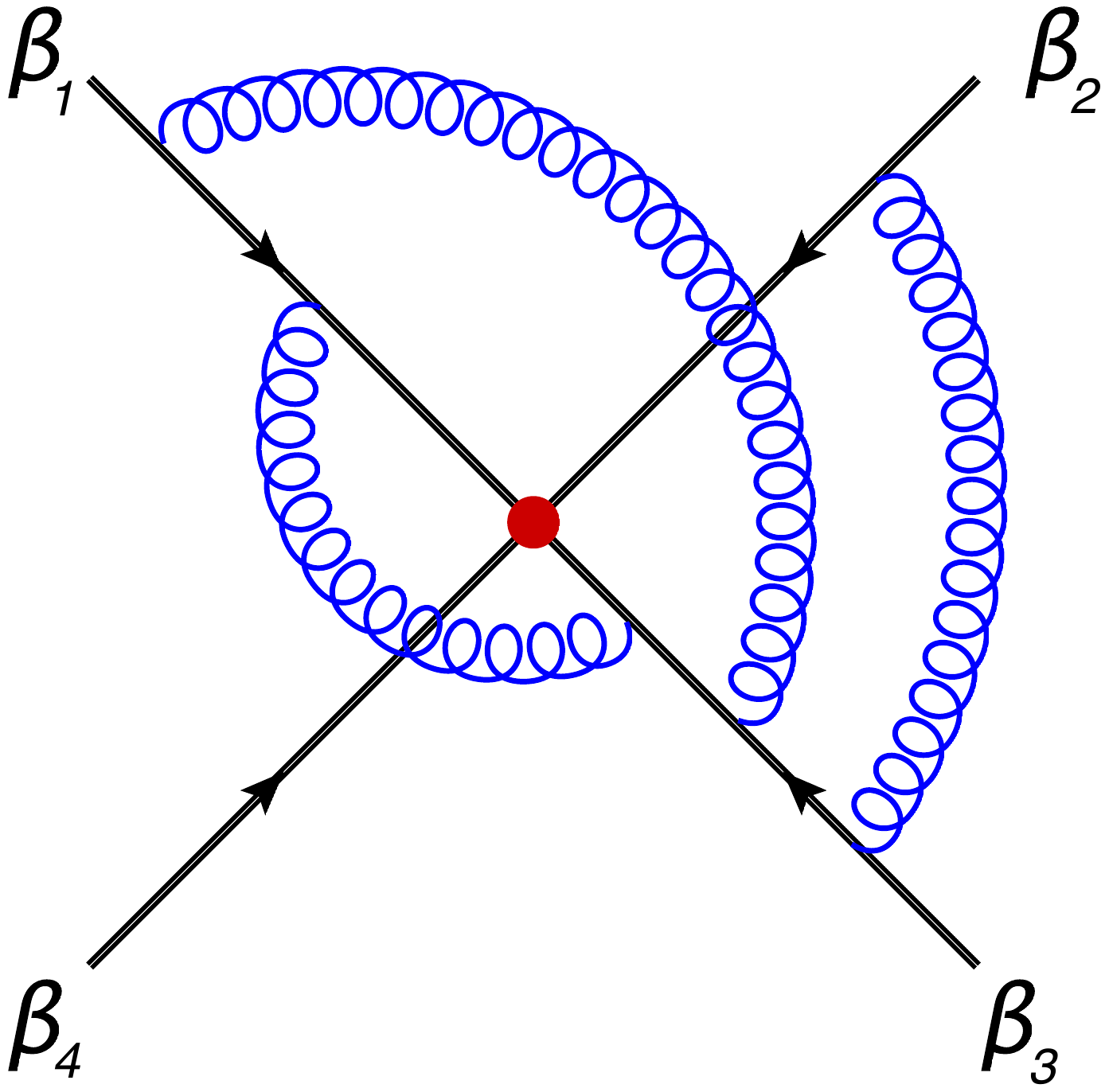}}\hspace*{10pt}
\scalebox{.25}{\includegraphics{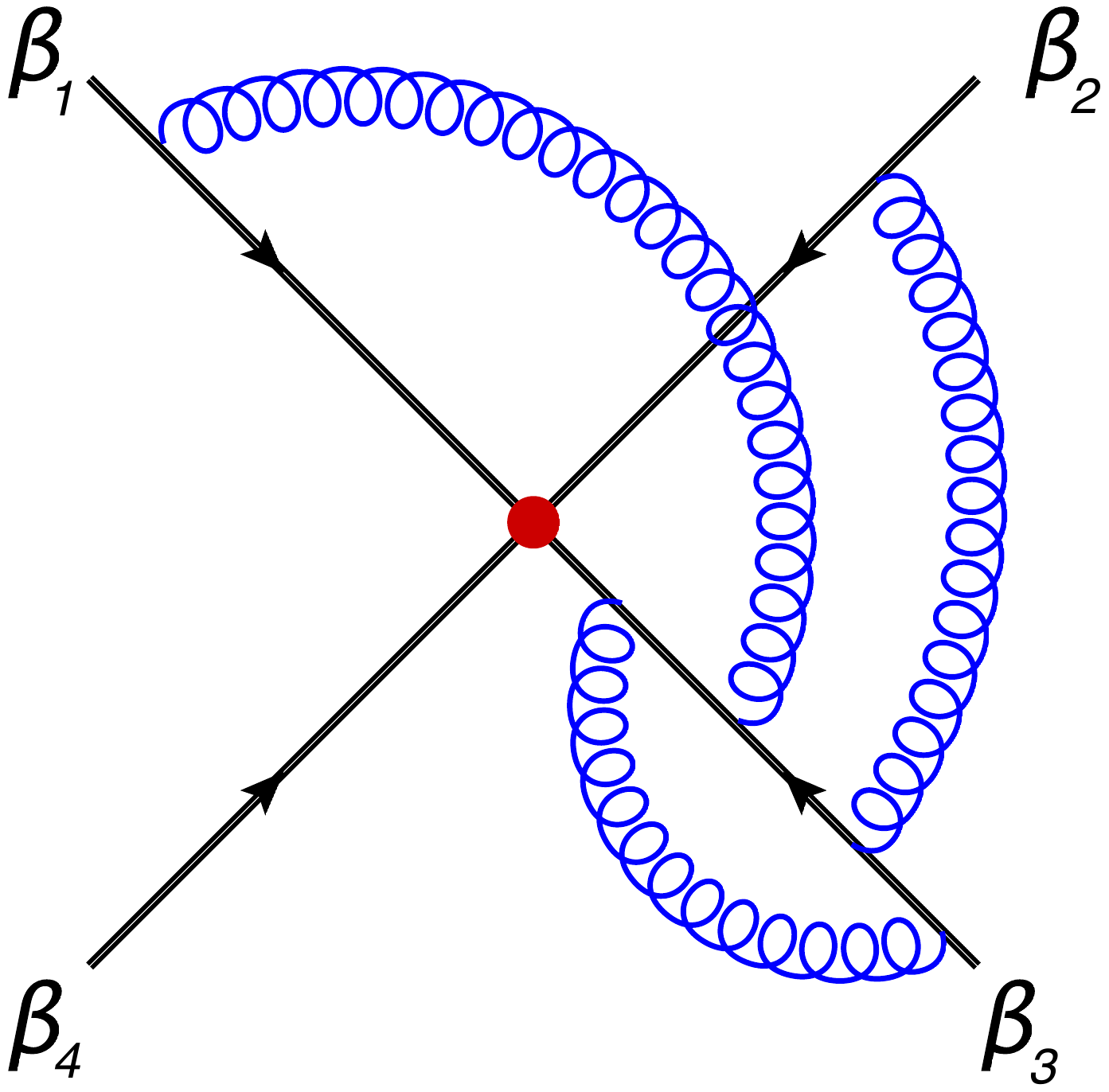}}\hspace*{10pt}
\caption{Representative 3-loop diagrams of webs connecting a subset of three out of the four Wilson lines.}
\label{4lines_3}
\end{center}
\end{figure*}

\begin{figure*}[!t]
\begin{center}
\scalebox{.25}{\includegraphics{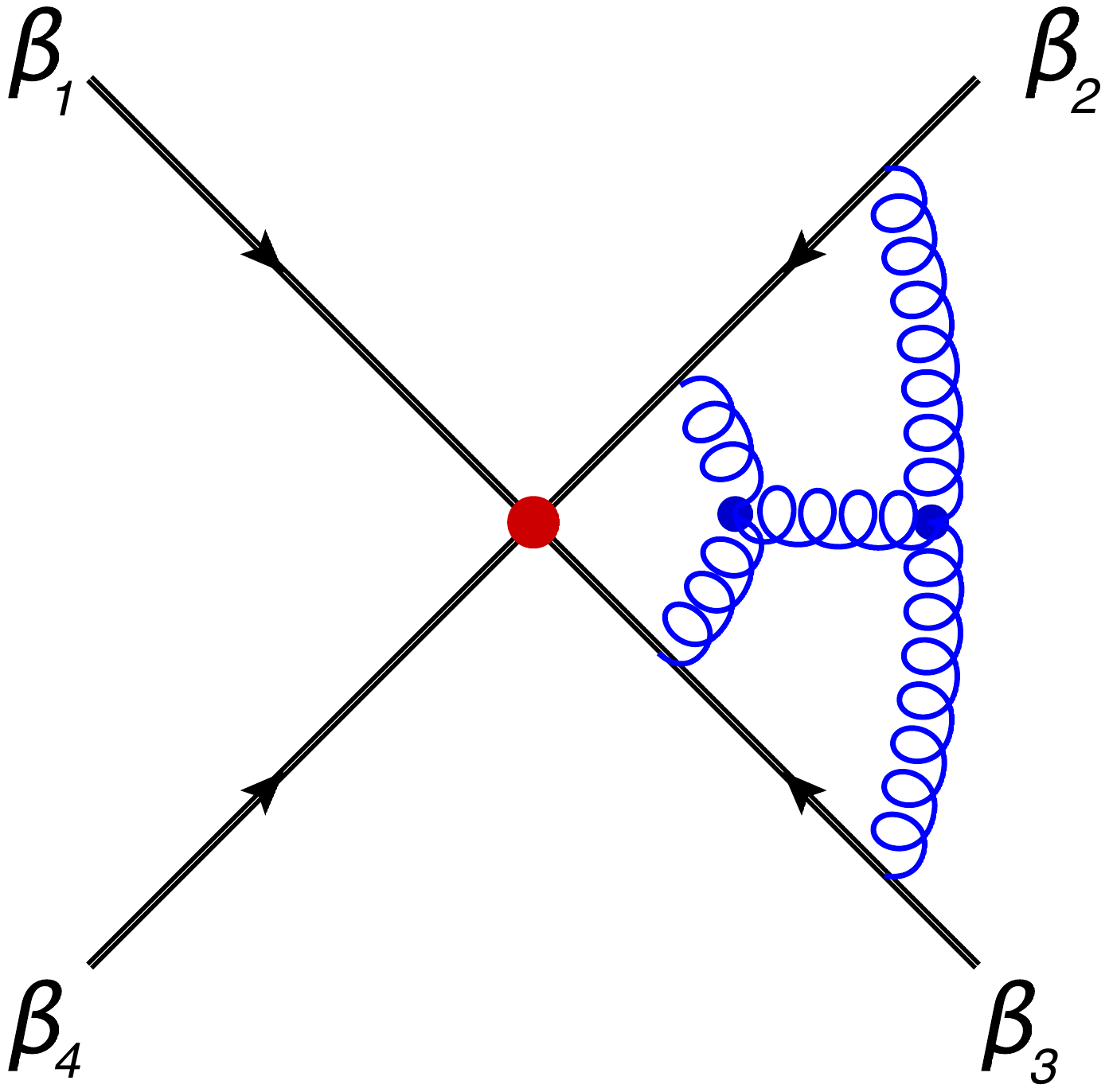}}\hspace*{10pt}
\scalebox{.25}{\includegraphics{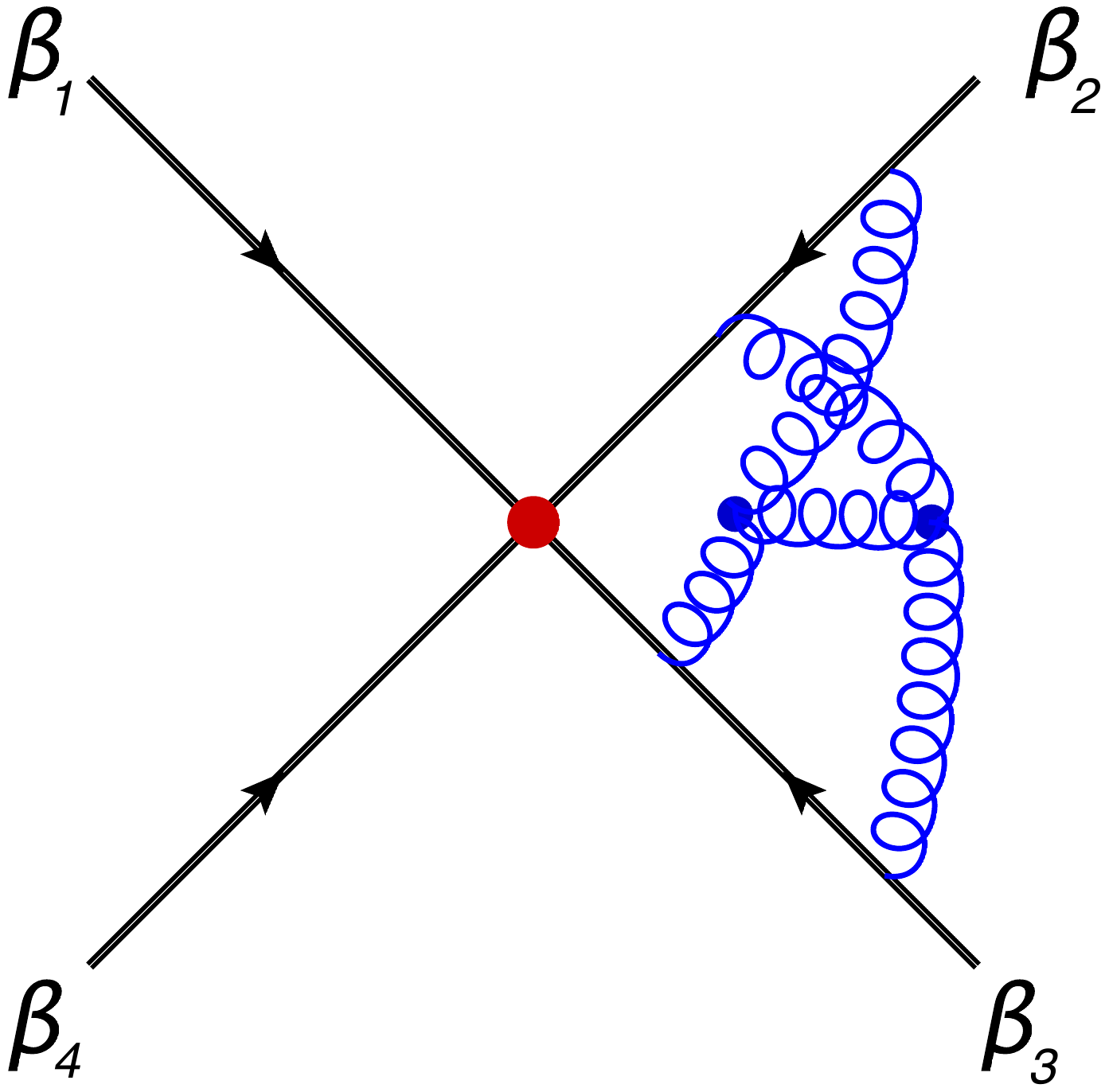}}\hspace*{10pt}
\scalebox{.25}{\includegraphics{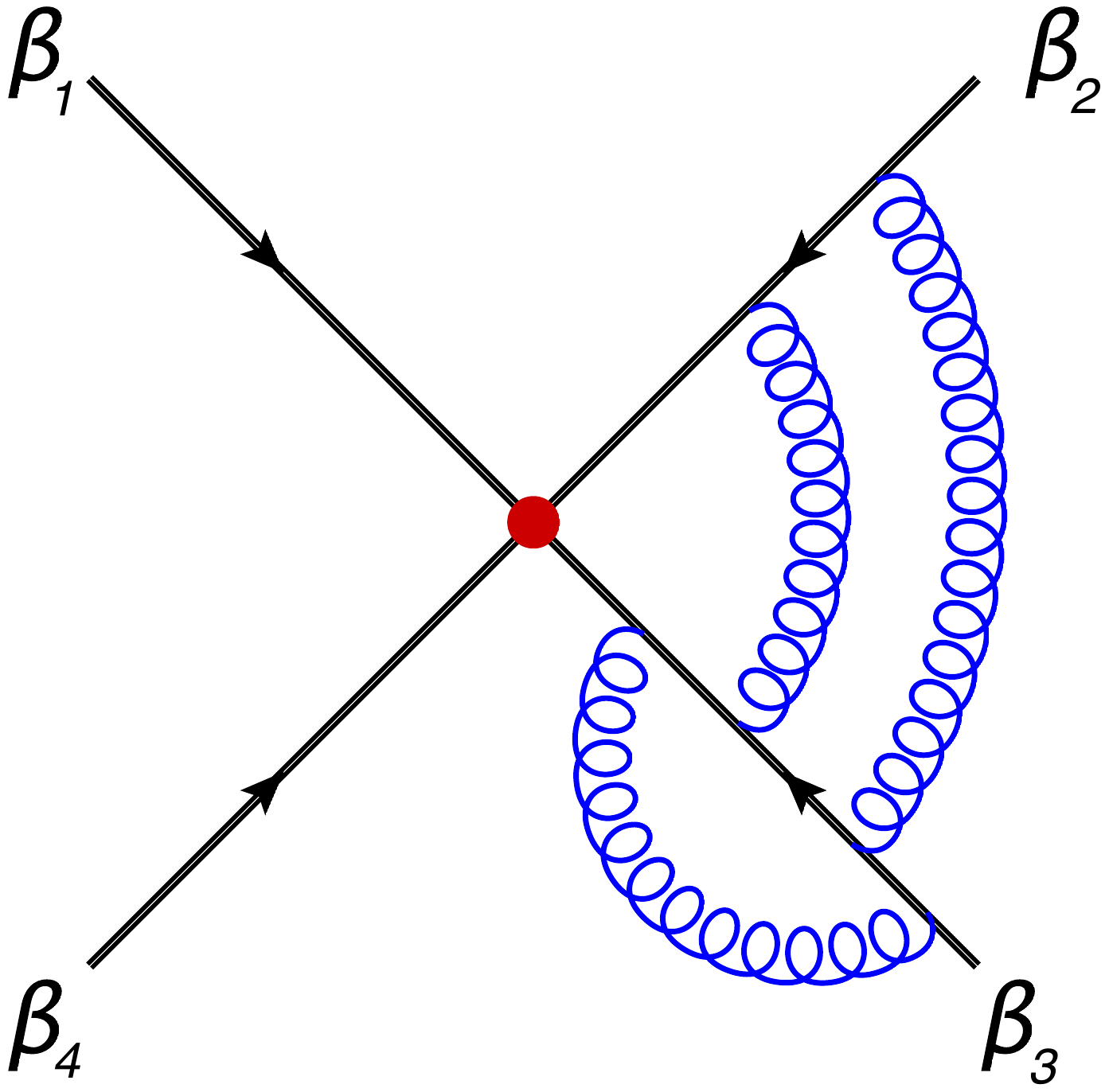}}\hspace*{10pt}
\scalebox{.25}{\includegraphics{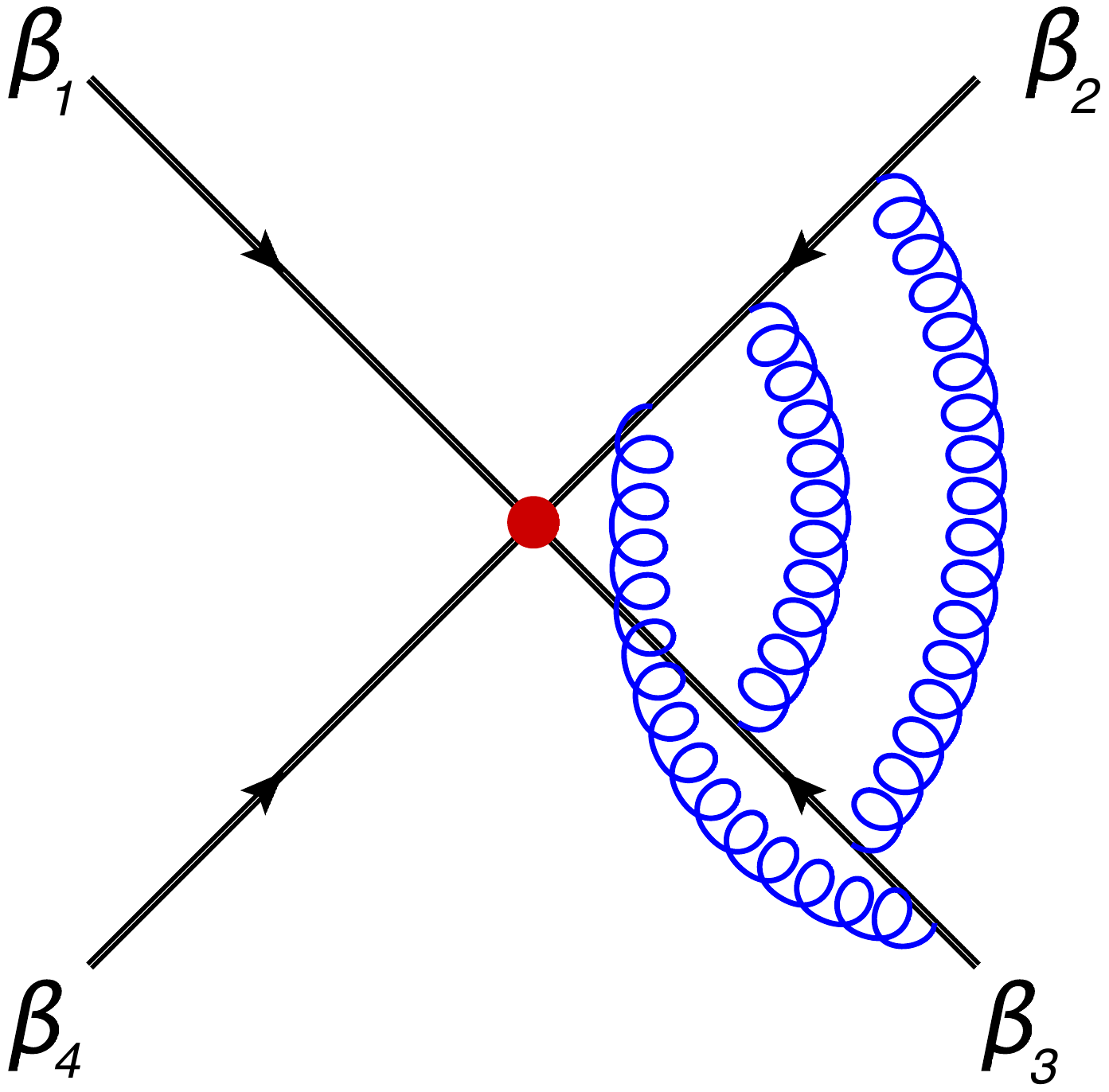}}
\caption{Representative 3-loop diagrams of webs connecting a subset of two out of the four Wilson lines.}
\label{4lines_2}
\end{center}
\end{figure*}

We note that $\Delta_n^{(3)}$ is independent of the details of the underlying theory and completely determined by soft gluon interactions. In particular, this implies that $\Delta_n^{(3)}$ is the same in QCD and in $\mathcal{N}=4$ Super Yang-Mills, and it is therefore expected to be a pure polylogarithmic function of weight five. Its functional form has been constrained by considering collinear limits and the Regge limit ~\cite{Gardi:2009qi,Dixon:2009gx,Becher:2009cu,Dixon:2009ur,Gardi:2009zv,Bret:2011xm,Ahrens:2012qz,Naculich:2013xa,Caron-Huot:2013fea}, but despite progress in understanding these limits it remained unclear whether three-loop corrections to the dipole formula are in fact present. The situation changed with the completion of the direct computation of $\Delta_n^{(3)}$ \cite{Almelid:2015jia} on which we report in the present talk.

\section{Computing connected graphs}

We set up the calculation of the soft anomalous dimension through the renormalization of a product of semi-infinite Wilson lines with four-velocities $\beta_k$, with $\beta_k^2\neq 0$. By considering non-lighlike lines we avoid collinear singularities, and obtain kinematic dependence via cusp angles $\gamma_{ij}\equiv 2\beta_i\cdot\beta_j/\sqrt{\beta_i^2\beta_j^2}$.
We eventually extract $\Delta_n^{(3)}$ for massless scattering by considering the asymptotic lightlike limit $\beta_k^2\to 0$, where the kinematic dependence reduces to CICRs as in Eq.~\eqref{eq:CICR}.

Considering the set of contributing diagrams at three loops, it is clear at the outset that the diagrams that connect the maximal number of Wilson lines, that is \emph{four} lines, shown in Fig.~\ref{4lines_connected}, have a special status: these are the only diagrams that depend on all six cusp angles $\gamma_{ij}$ with  \hbox{$1\leq i<j\leq 4$}. Hence these four diagrams are expected to involve non-trivial dependence on CICRs (defined in Eq.~(\ref{eq:CICR})). Importantly, this kinematic dependence remains in place upon taking the simultaneous lightlike limit, $\gamma_{ij}\to -\infty$.
In contrast, all other webs reduce in this limit to a sum of products of logarithms of $\gamma_{ij}$. This applies in particular to the webs of Fig.~\ref{4lines_non-connected}: these webs connect all of the four lines, but they never involve any set of four angles that may form a cross ratio as in Eq.~(\ref{eq:CICR}). It is clear that webs connecting three or two lines out of the four, as in Figs. \ref{4lines_3} 
and~\ref{4lines_2}, cannot give rise to cross ratios, and so they reduce to polynomials in logarithms of $\gamma_{ij}$ for near lightlike kinematics. Of course, cross ratios may be formed upon summing the webs of Figs.~\ref{4lines_non-connected}, \ref{4lines_3} and~\ref{4lines_2}, but these contributions are necessarily polynomial in logarithms of the CICRs.

It follows that the primary ingredient in deriving $\Delta_4^{(3)}\left(\left\{\rho_{ijkl}\right\}\right)$ is the computation of the four-line connected diagrams in Fig.~\ref{4lines_connected}. Below we briefly describe the strategy of the calculation and the result we obtain for these diagrams, before presenting the complete result for the anomalous dimension. The computation of all diagrams will be discussed in dedicated publication~\cite{longpaper}. 

We set up the calculation in configuration space, with four non-lightlike Wilson lines with four-velocities $\beta_k$. The position of the three- and four-gluon vertices off the Wilson lines are integrated over in $D=4-2\epsilon$ dimensions. Following Ref.~\cite{Gardi:2011yz,Gardi:2013saa}, we introduce an infrared regulator which exponentially suppresses contributions far along the Wilson lines. This is necessary to capture the ultraviolet singularity associated with the renormalization of the vertex where the Wilson lines meet. 
Upon performing the integral over the overall scale, we observe that each of the diagrams in Fig.~\ref{4lines_connected} has a single $1/\epsilon$ ultraviolet pole, without any subdivergences. The contribution of each diagram to the soft anomalous dimension is the coefficient of that pole, which is finite in $D=4$ dimensions.

Next, considering the leftmost diagram in Fig.~\ref{4lines_connected}, we observe that for fixed gluon-emission vertices along the Wilson lines, 
the integral over the position of the four-gluon vertex gives rise to a four-mass one-loop box integral in 4 dimensions; Similarly, in each of the remaining three diagrams in Fig.~\ref{4lines_connected}, the integrals over the positions of the two three-gluon vertices  yield a four-mass diagonal-box two-loop integral\footnote{Some more details on this computation were presented in the previous Loops and legs conference \cite{Gardi:2014kpa}.}.  
We proceed by deriving multifold Mellin-Barnes (MB) representations for each of these off-shell four-point functions.
 
Next we integrate over the position of the gluon emission vertices along the Wilson lines, obtaining a MB representation of each of the connected graphs for the general non-lightlike case, depending on all of the six cusp angles~$\left\{\gamma_{ij}\right\}$. We proceed by applying standard techniques~\cite{Czakon:2005rk} to perform a simultaneous asymptotic expansion near the lightlike limit $\gamma_{ij}\to -\infty$, where we neglect any term suppressed by powers of $1/\gamma_{ij}$, obtaining a sum of lower-dimensional MB integrals. These  are converted into parametric integrals using the methods of Ref.~\cite{Anastasiou:2013srw}, which we then performed by means of modern analytic integration techniques~\cite{Goncharov.A.B.:2009tja}. The  result for the leftmost diagram in Fig.~\ref{4lines_connected} reads:
\begin{align}
w_{4g}=&\frac{1}{\epsilon}
\left(\frac{\alpha_s}{4\pi}\right)^3{\rm \bf T}_1^a{\rm \bf T}_2^b{\rm \bf T}_3^c{\rm \bf T}_4^d\Big[ f^{abe}  f^{cde} \frac{z\bar{z}-z-\bar{z}}{z-\bar{z}}
+ f^{ade}  f^{bce} \frac{1-z\bar{z}}{z-\bar{z}}
+ f^{ace}f^{bde}\frac{1-z-\bar{z}}{z-\bar{z}}
\Big]\, g_1(z,\bar{z},\left\{\gamma_{ij}\right\})
\end{align}
and the one for the second diagram takes the form
\begin{align}
w_{(12)(34)}=&\frac{1}{\epsilon}
\left(\frac{\alpha_s}{4\pi}\right)^3 {\rm \bf T}_1^a{\rm \bf T}_2^b{\rm \bf T}_3^c{\rm \bf T}_4^d\,\, f^{abe}  f^{cde} \Big[g_0(z,\bar{z},\left\{\gamma_{ij}\right\})  - \frac{z\bar{z}-z-\bar{z}}{z-\bar{z}}  g_1(z,\bar{z},\left\{\gamma_{ij}\right\})\Big]\,,
\end{align}
where $g_0$ and $g_1$ are pure polylogarithmic functions of uniform weight five in the variables 
$z\equiv z_{ijkl}$ and $\bar{z}\equiv \bar{z}_{ijkl}$ which are related to the CICRs of Eq.~(\ref{eq:CICR}) via
\beq
\label{zbarzdef}
z_{ijkl}\,\bar{z}_{ijkl} = \rho_{ijkl} {\rm~~and~~} (1-z_{ijkl})\,(1-\bar{z}_{ijkl})=\rho_{ilkj}\,.
\eeq
The remaining two diagrams in Fig.~\ref{4lines_connected} can be obtained from $w_{(12)(34)}$ by appropriate permutations of the lines. 
The sum over all four connected graphs,
$w_{\rm con.}=w_{4g}+w_{(12)(34)}+w_{(13)(24)}+w_{(14)(23)}$,
 displays a drastic simplification as compared to individual diagrams, namely, individual graphs are not pure functions but the sum is. Specifically, the function $g_1$, which appears in all of them, exactly cancels in the sum, and one is left with three permutations of the function $g_0$, which has no rational prefactor. This is in agreement with the expectation that (maximally helicity-violating) amplitudes in $\mathcal{N}=4$ Super Yang-Mills are pure and have a uniform maximal weight. 

The next simplification occurs upon applying the Jacobi identity to the sum of connected 4-line webs:
\begin{align}
\begin{split}
w_{\rm con.}=\frac{1}{\epsilon}
\left(\frac{\alpha_s}{4\pi}\right)^3 {\rm \bf T}_1^a{\rm \bf T}_2^b{\rm \bf T}_3^c{\rm \bf T}_4^d\,\,  \bigg[ f^{abe}  f^{cde}
{\cal G}_1(z,\bar{z},\left\{\gamma_{ij}\right\})
+f^{ade}  f^{bce}{\cal G}_2(z,\bar{z},\left\{\gamma_{ij}\right\})
\bigg]\,,
\end{split}
\end{align}
where 
$
{\cal G}_1=g_0
+\left.\left[g_0 \right]\right\vert_{j\leftrightarrow k}
$
and 
$
{\cal G}_2=
\left.\left[g_0 \right]\right\vert_{j\to l\to k\to j }
+\left.\left[g_0 \right]\right\vert_{j\leftrightarrow k}
$.
Crucially, the functions ${\cal G}_{1,2}(z,\bar{z},\left\{\gamma_{ij}\right\})$ separate as follows:
\begin{equation}
\label{separ}
{\cal G}_{1,2}(z,\bar{z},\left\{\gamma_{ij}\right\})
=P_{1,2}(z,\bar{z}) +Q_{1,2}\left(\left\{\log \left(\gamma_{ij}\right)\right\}\right)\,,
\end{equation}
where $P_{1,2}(z,\bar{z})$ is a sum of harmonic polylogarithms (of weight 5) depending exclusively of on CICRs via $z$ and $\bar{z}$,
while $Q_{1,2}\left(\left\{\log \left(\gamma_{ij}\right)\right\}\right)$ is a polynomial in the logarithms of $\gamma_{ij}$. This split must have happened for the full result for $\Delta_n^{(3)}$ to be a function of CICRs: indeed $Q_{1,2}\left(\left\{\log \left(\gamma_{ij}\right)\right\}\right)$ 
cancels against contributions of the remaining diagrams\footnote{One notes that the separation in (\ref{separ}), while highly constraining, is not unique: powers of logarithms of CICRs can be expressed in either way. The computation of the remaining diagrams of Figs.~\ref{4lines_non-connected}, \ref{4lines_3} and~\ref{4lines_2}, uniquely fixes the answer.} -- which are also polynomial in $\left\{\log \left(\gamma_{ij}\right)\right\}$ -- leaving behind pure CICR dependence.

\section{Colour structure and colour conservation at three loops}

Let us now turn to discuss the colour structure of the soft anomalous dimension for $n$ coloured lines.
According to the non-Abelian exponentiation theorem~\cite{Gardi:2013ita} the colour factors in $\Delta_n$ must all correspond to connected graphs\footnote{In this context a ``connected graph'' is one that remains connected upon removing all Wilson lines, so for example all diagrams in Fig.~\ref{4lines_connected} are connected while all those of Fig.~\ref{4lines_non-connected} are non-connected.
This does not imply that the latter do not contribute -- they do, but with the colour factors of the former.
 For further details see~Refs.~\cite{Gardi:2010rn,Mitov:2010rp,Gardi:2011wa,Gardi:2011yz,Gardi:2013ita,Gardi:2013saa}.}.
Thus, at three loops we expect the ``quadrupole'' colour structures of  Fig.~\ref{4lines_connected}, i.e., ${\rm \bf T}_i^a{\rm \bf T}_j^b{\rm \bf T}_k^c{\rm \bf T}_l^d f^{abe}  f^{cde}$ plus permutations, where the four lines connected ($i,j,k$ and $l$) are any subset of four out of the $n$ lines.

The next question is then whether any other colour factor is admissible in $\Delta_n^{(3)}$, namely ones that involve fewer than four lines. 
One possibility could be tripole corrections correlating three partons, with colour factors proportional to 
 ${\rm i}f^{abc} {\rm \bf T}_i^a{\rm \bf T}_j^b{\rm \bf T}_k^c$. Such tripoles appear starting from two loops for non-lightlike Wilson lines~\cite{Kidonakis:2009ev,Mitov:2009sv,Becher:2009kw,Beneke:2009rj,Czakon:2009zw,Ferroglia:2009ep,Ferroglia:2009ii,Chiu:2009mg,Mitov:2010xw,Gardi:2013saa,Henn:2013tua},  but are excluded in the lightlike case at any order because the corresponding kinematic dependence on the three momenta is bound to violate the rescaling symmetry constraints~\cite{Gardi:2009qi,Becher:2009cu,Gardi:2009zv}.
While a constant correction proportional to ${\rm i}f^{abc} {\rm \bf T}_i^a{\rm \bf T}_j^b{\rm \bf T}_k^c$ is excluded by Bose symmetry, kinematic-independent corrections involving three lines of the form  $f^{abe}f^{cde}\left\{{\rm \bf T}_i^a,  {\rm \bf T}_i^d\right\}   {\rm \bf T}_j^b {\rm \bf T}_k^c$ as the first diagram on 
Fig.~\ref{4lines_3}, are admissible and do indeed appear.   

We conclude that the general form of the non-dipole correction to the soft anomalous dimension for $n$ coloured lines is given by 
\begin{align}\label{eq:expected_Delta}
 &\Delta_n^{(3)}\left(\left\{\rho_{ijkl}\right\}\right) = 16\,f_{abe}f_{cde} 
\Big\{-C\, \sum_{i=1}^n\sum_{\substack{{1\leq j<k\leq n}\\ j,k\neq i}}\left\{{\rm \bf T}_i^a,  {\rm \bf T}_i^d\right\}   {\rm \bf T}_j^b {\rm \bf T}_k^c  \quad +
\\
\nonumber
& 
{\sum_{1\leq i<j<k<l\leq n}}
\Big[
 {\rm \bf T}_i^a  {\rm \bf T}_j^b   {\rm \bf T}_k^c {\rm \bf T}_l^d   \, {\cal F}(\rho_{ikjl},\rho_{iljk}) 
+{\rm \bf T}_i^a  {\rm \bf T}_k^b {\rm \bf T}_j^c   {\rm \bf T}_l^d    
\, {\cal F}(\rho_{ijkl},\rho_{ilkj}) 
+ {\rm \bf T}_i^a   {\rm \bf T}_l^b  {\rm \bf T}_j^c    {\rm \bf T}_k^d 
\, {\cal F}(\rho_{ijlk},\rho_{iklj}) \Big]
\Big\}\,,
\end{align}
where $C$ is a constant and ${\cal F}$ is a function of two CICRs. Note that the contribution proportional to the constant $C$ is present starting from the three-line case, $n=3$.
Both $C$ and $\mathcal{F}$ are independent of the colour degrees of freedom. The terms in this sum are not all independent, because of the antisymmetry of the structure constants and the Jacobi identity. 
We emphasise that $C$ and $\mathcal{F}$ are independent of the number of legs~$n$. We can therefore determine these functions by considering the simplest case of four Wilson lines, $\Delta_4^{(3)}$. 

In organising the calculation we made use of non-Abelian exponentiation, and computed \emph{webs}, namely diagrams that contribute directly to the exponent.  A web can be either an individual connected diagram, as in Fig.~\ref{4lines_connected}, or a set of non-connected diagrams which are related by permuting the order of gluon attachments to the Wilson lines~\cite{Gardi:2010rn,Mitov:2010rp,Gardi:2011wa,Gardi:2011yz,Gardi:2013ita}; representative diagrams from such webs are shown in Fig.~\ref{4lines_non-connected}. In either of these cases, the contribution to $\Delta_4^{(3)}$ is associated with fully connected colour factors. The classification of webs connecting four and three Wilson lines was done in Ref.~\cite{Gardi:2013ita}. 

Another important element in organising the calculation is colour conservation. The anomalous dimension $\Gamma_n$ is an operator in colour space that acts on the hard amplitude, which is a colour singlet and must therefore satisfy~\cite{Catasterm}
\begin{align}
\begin{split}
\left(\sum_{i=1}^n{\rm \bf T}_i^a\right)\mathcal{H}_n = 0\,.
\end{split}
\end{align}
This colour conservation constraint is implicit in Eqs. (\ref{dipole_formula}) and (\ref{eq:expected_Delta}).
When computing $\Delta_4^{(3)}$ one may form a colour basis by systematically eliminating ${\rm \bf T}_4$ in favour of ${\rm \bf T}_i$, $1\le i\le3$, thereby reducing all four-line colour factors to three-line ones. 
This way colour conservation relates between diagrams connecting a different number of Wilson lines: the diagrams in Figs. \ref{4lines_3} and~\ref{4lines_2}, which connect three or two Wilson lines, contribute together with those connecting four lines. Let us see this explicitly.
The sum of all three-loop webs connecting four lines can be cast into the Bose symmetric form
\begin{align}
\label{G4}
\begin{split}
G_4(1,2,3,4) = {\rm \bf T}_1^a {\rm \bf T}_2^b {\rm \bf T}_3^c {\rm \bf T}_4^d\, \Big[&f^{abe}f^{cde} H_4[(1,2),(3,4)]\,+
\\&  
f^{ace}f^{bde} H_4[(1,3),(2,4)] + f^{ade}f^{bce} H_4[(1,4),(2,3)]\Big]\,,
\end{split}
\end{align}
where the kinematic function $H_4$ satisfies the following permutation properties: $H_4[(1,2),(3,4)]= -H_4[(2,1),(3,4)]= H_4[(3,4),(1,2)]$;
this function depends on logarithms of cusp angles as well as on non-trivial functions of CICRs. Using colour conservation to eliminate ${\rm \bf T}_4$ in favour of the sum of the other three generators, we convert the result to a three-line colour basis:
\begin{align}
\label{G_4}
G_4(1,2,3,4) = -\frac12 f^{abe}f^{cde}\!\!\!\! \sum_{\substack{{(i,j,k)\in (1,2,3)}\\j<k}}\!\!\! \left\{{\rm \bf T}_i^a,  {\rm \bf T}_i^d\right\}   {\rm \bf T}_j^b {\rm \bf T}_k^c   \,
\bigg(H_4[(i,j),(k,4)]+H_4[(i,k),(j,4)]\bigg)\,.
\end{align}

Let us consider next diagrams that connect fewer Wilson lines. The sum of all two-line three-loop diagrams may be written as
\begin{align}
G_2(1,2)\,={\bf dipole}\,-\, f^{abe}f^{cde} 
\left\{ {\rm \bf T}_1^a, {\rm \bf T}_1^d\right\} \Big\{{\rm \bf T}_2^b, {\rm \bf T}_2^c\Big\} \, H_2(1,2)\,,
\end{align}   
where the first term represents the dipole ${\bf T}_1\cdot {\bf T}_2$ contribution to $\Gamma_{n}^{\rm dip.}$ of Eq.~(\ref{dipole_formula}). In contrast, the second term involving an anti-commutator on each of the lines is relevant for the calculation of~$\Delta_n^{(3)}$; its kinematic dependence is contained in $H_2(1,2)=H_2(2,1)$. Similarly, the sum of all three-line diagrams takes the form
\begin{align}
G_3(1,2,3)= f^{abe}f^{cde}\!\!\!\!\!\!\!\!\sum_{\substack{(i,j,k)\in (1,2,3)\\j<k}} \!\!\!\!\!\!\!\left\{ {\rm \bf T}_i^a, {\rm \bf T}_i^d\right\} {\rm \bf T}_j^b {\rm \bf T}_k^c \, H_3[i,\{j,k\}] \,,
\end{align}  
with $H_3[i,\{j,k\}] = H_3[i,\{k,j\}]$. We omitted here the tripole term, proportional to $f^{abc} {\rm \bf T}_1^a {\rm \bf T}_2^b {\rm \bf T}_3^c$, which vanishes for lightlike kinematics where $\gamma_{ij}\to -\infty$. Note that in this limit $H_2$ and $H_3$ are necessarily polynomials in $\log(-\gamma_{ij})$. 

Summing over all subsets of two and three lines out of four and using colour conservation, we have
\begin{eqnarray}
&&G_2(1,2,3,4)+ G_3(1,2,3,4) =\, \text{\bf dipoles}\,+\, f^{abe}f^{cde}\, \bigg[\sum_{\small \substack{{(i,j,k)\in (1,2,3)}\\j<k}}\left\{{\rm \bf T}_i^a,  {\rm \bf T}_i^d\right\}   {\rm \bf T}_j^b {\rm \bf T}_k^c   \,
\,\overline{U}(i,\{j,k\},4)\nonumber \\&&\hspace*{40pt}-\, \frac12 \sum_{1\leq i\leq j\leq 3}  \left\{{\rm \bf T}_i^a,  {\rm \bf T}_i^d\right\}  
\left\{ {\rm \bf T}_j^b, {\rm \bf T}_j^c\right\} \bigg(\overline{H}_3[i,\{j,4\}] +\overline{H}_3[j,\{4,i\}] +\overline{H}_3[4,\{j,i\}]\bigg)
\bigg] 
\label{G23}
\end{eqnarray}
where
\begin{align}
\begin{split}
\overline{U}(i,\{j,k\},4)\equiv\! \overline{H}_3[i,\{j,k\}]\! -\!\overline{H}_3[i,\{j,4\}]\! -\!\overline{H}_3[i,\{k,4\}]\! -\!\overline{H}_3[4,\{i,j\}] \!-\!\overline{H}_3[4,\{i,k\}]\!+\!\overline{H}_3[4,\{j,k\}]  
\end{split}
\end{align}
with $\overline{H}_3[i,\{j,k\}]\equiv H_3[i,\{j,k\}] +H_2(i,j) +H_2(i,k)$.

The three- and two-line contributions of Eq.~(\ref{G23}) must be added to the contribution of the four-line diagrams in Eq.~(\ref{G_4}) to obtain the final, gauge-invariant result for the anomalous dimension, $\Delta_4^{(3)}=G_4(1,2,3,4)+G_3(1,2,3,4)+G_2(1,2,3,4)$. This may then be contrasted with the general form for $\Delta_4^{(3)}$ in Eq.~(\ref{eq:expected_Delta}). Upon applying colour conservation to the latter, the comparison leads to the following conclusions:
\begin{itemize}
\item{} The combination multiplying the two-line colour factor in Eq.~(\ref{G23}) must be proportional to the constant $C$ in Eq.~(\ref{eq:expected_Delta}):
\begin{align}
\label{C_in_terms_of_H3}
\!\!C= \frac13\Big(
 \overline{H}_3[i,\{j,k\}] 
+\overline{H}_3[j,\{k,i\}] 
+\overline{H}_3[k,\{j,i\}]
\Big)\,,
\end{align}   
\item{}
The function $\mathcal{F}$ is obtained through the following combination of four-, three- and two-line kinematic functions $H_n$:
\begin{align}
\begin{split}
\label{calF_in_terms_of_H}
&\!\!\mathcal{F}(\rho_{ijkl},\rho_{ilkj})= H_4[(i,j),(k,l)]
-\frac23 \Big(
 \overline{H}_3[i,\{j,k\}]
-\overline{H}_3[i,\{j,l\}]
-\overline{H}_3[j,\{i,k\}]
+\overline{H}_3[j,\{i,l\}]
\\
&\hspace*{140pt}
+\overline{H}_3[k,\{i,l\}]
-\overline{H}_3[k,\{j,l\}]
-\overline{H}_3[l,\{i,k\}]
+\overline{H}_3[l,\{j,k\}]\Big)\,.
\end{split}
\end{align}
\end{itemize}
The above equations put strong constraints on the kinematic functions $H_n$: the function $\mathcal{F}$ depends on CICRs, while the individual functions $H_n$ on the right-hand side of Eq.~(\ref{calF_in_terms_of_H}) depend on logarithms of cusp angles. These must therefore conspire to combine into logarithms of CICRs. In addition, $C$ is a constant, so the kinematic dependence of the functions $\overline{H}_3$ must cancel in the sum in Eq.~\eqref{C_in_terms_of_H3}. Our computation satisfies all these constraints, providing a strong check of the result.

\section{The three-loop correction to the soft anomalous dimension}

Adding up all contributing webs according to Eqs.~(\ref{calF_in_terms_of_H}) and \eqref{C_in_terms_of_H3}, we find the following results for the function $\mathcal{F}$ and the constant $C$ of Eq.~(\ref{eq:expected_Delta}):
\begin{align}\begin{split}
\mathcal{F}(\rho_{ijkl},\rho_{ilkj}) &\, = F(1-z_{ijkl}) - F(z_{ijkl})\,,\\
C &\,= \zeta_5 + 2\zeta_2\,\zeta_3\,,
\end{split}\end{align}
where we recall that $z=z_{ijkl}$ and $\bar{z}=\bar{z}_{ijkl}$ are related to the CICRs by Eq.~(\ref{zbarzdef}) and 
\begin{equation}
F(z)\, = \mathcal{L}_{10101}(z) +2\,\zeta_2\,\left[\mathcal{L}_{001}(z)+\mathcal{L}_{100}(z)\right]\,,
\label{eq:F_definition}
\end{equation}
where the functions $\mathcal{L}_{w}(z)$ are Brown's \emph{single-valued harmonic polylogarithms} (SVHPLs)~\cite{BrownSVHPLs} (see also Ref.~\cite{Dixon:2012yy}), where $w$ is a word made out of 0's and 1's. Note that we kept implicit the dependence of these functions on~$\bar{z}$. SVHPLs can be expressed in terms of ordinary harmonic polylogarithms (HPLs)~\cite{Remiddi:1999ew} in $z$ and $\bar{z}$. The result for $F$ in terms of HPLs is attached in computer-readable format to Ref.~\cite{Almelid:2015jia}.

Let us now briefly discuss the main features of the result. First, we note that
while $F(z)$ is defined everywhere in the physical parameter space, it is only single-valued in the part of the Euclidean region (the region where all invariants are spacelike, $p_i\cdot p_j <0$) where $z$ and $\bar{z}$ are complex conjugate to each other. Single-valuedness ensures that $\Delta_n^{(3)}$ has the correct branch cut structure of a physical scattering amplitude~\cite{Gaiotto:2011dt,Dixon:2012yy}: it is possible to analytically continue the function to the entire Euclidean region while the function remains real throughout~\cite{Chavez:2012kn}. Next note that if one considers $F(z)$ as a function of two independent variables $z$ and $\bar{z}$ (not a complex conjugate pair) this function has branch points for $z$ and $\bar{z}$ at $0$, $1$ and $\infty$. Crossing momenta from the final to the initial state is realized by taking monodromies around these points.

Making the permutation (Bose) symmetry manifest, the final answer may be written as:
\begin{align}
\label{final_result_n}
\begin{split}
\Delta_n^{(3)}&=16\!\!\!\!{\sum_{1\leq i<j<k<l\leq n}}\!\!\!\!\! {\bf T}_i^a {\bf T}_j^b {\bf T}_k^c {\bf T}_l^d \,\Bigg[
  f_{abe}f_{cde} \Big( F(1-1/z_{ijkl},1-1/\bar{z}_{ijkl}) - F(1/z_{ijkl},1/\bar{z}_{ijkl}) \Big) \\
&\hspace*{40pt}+f_{ace}f_{bde} \Big( F(1-z_{ijkl},1-\bar{z}_{ijkl}) - F(z_{ijkl},\bar{z}_{ijkl}) \Big)
\\&\hspace*{40pt}
+f_{ade}f_{bce} \Big( F(1/(1-z_{ijkl}), 1/(1-\bar{z}_{ijkl})) - F(z_{ijkl}/(z_{ijkl}-1),\bar{z}_{ijkl}/(\bar{z}_{ijkl}-1) \Big)\Bigg]
\\
&\hspace*{40pt}
- C \,\,\sum_{i=1}^n \sum_{\substack{{1\leq j<k\leq n}\\ \,j,k\neq i}}  f_{abe} f_{cde} \left\{{\bf T}_i^a,{\bf T}_i^b\right\} {\bf T}_j^b {\bf T}_k^c
\end{split}
\end{align}
where, as in Eq.~(\ref{eq:expected_Delta}), colour conservation among the $n$ lines is implicit.
For any subset of four lines ($i,j,k$ and $l$) Bose symmetry is realised on the function $F$ by the action of the group $S_3$ which keeps the momentum $p_i$ fixed and permutes the remaining three momenta. As is clear from Eq.~(\ref{final_result_n}), this group acts on the space of SVHPLs by change of arguments generated by the transformations $(z,\bar{z})\mapsto (1-\bar{z}, 1-z)$ and $(z,\bar{z})\mapsto (1/\bar{z}, 1/z)$, with $z\equiv z_{ijkl}$. Geometrically this corresponds to exchanging the three singularities at $z\in\{0,1,\infty\}$. Moreover, the space of all HPLs, and hence also SVHPLs, is closed under the action of this $S_3$; this gives rise to functional relations among HPLs with different arguments, making it possible to express all the terms in Eq.~\eqref{final_result_n} through SVHPLs with argument~$z$. 

An additional symmetry group $\mathbb{Z}_2$ arises from the definition of $(z,\bar{z})$ in Eq.~(\ref{zbarzdef}), which is invariant under swapping the two, $z\leftrightarrow \bar{z}$. Hence $F(z)$ must be invariant under this transformation, i.e. $F(\bar{z}) = F(z)$. This symmetry is realised on the space of SVHPLs by the operation of reversal of words, namely, if $w$ is a word made out of 0's and 1's, and $\widetilde{w}$ the reversed word, then we have $\mathcal{L}_w(\bar{z}) = \mathcal{L}_{\widetilde{w}}(z) +\ldots$, where the dots indicate terms proportional to  multiple zeta values. Even functions then correspond to `palindromic' words (possibly up to multiple zeta values), and indeed Eq.~\eqref{eq:F_definition} is `palindromic'.

Finally, let us comment on the momentum conserving limit of $\Delta_4^{(3)}$, which corresponds to two-to-two massless scattering. In this limit we have $\bar{z} = z = s_{12}/s_{13} = -s/(s+t)$. It follows that for two-to-two massless  scattering $F(z)$ can be expressed entirely in terms of HPLs with indices $0$ and $-1$ depending on $s/t$, in agreement with known results for on-shell three-loop four-point integrals~\cite{Smirnov:2003vi,Bern:2005iz,Henn:2013tua}.

A further consistency check of the result is available upon specialising to the Regge limit\footnote{Taking the Regge limit requires analytic continuation to the physical region of $2\to 2$ scattering, to be discussed in detail in~\cite{longpaper}.}. By  expanding Eq.~\eqref{eq:F_definition} at large $s/(-t)$ we find no $\alpha_s^3\, \ln^p\left({s}/({-t})\right)$ for any~$p>0$: $\Delta_4^{(3)}$ simply tends to a constant in this limit. This is entirely consistent with the behaviour of a two-to-two scattering amplitude in the Regge limit~\cite{Bret:2011xm,Caron-Huot:2013fea,Caron-Huot:TBA}; indeed, the dipole formula alone is consistent with predictions from the Regge limit through next-to-next-to-leading logarithms at three loops~\cite{Caron-Huot:TBA}.

\section{Two-particle collinear limits}

Finally, let us comment on the behaviour of $\Delta_n^{(3)}$ in the limit where two final-state partons become collinear. 
A well-known property of an $n$-parton scattering amplitude is that the limit where any two coloured partons become collinear can be related to an $(n-1)$-parton amplitude: 
\begin{align} 
\label{Sp_M}
\!\!{\cal M}_n \left(p_1,p_2,\{p_j\}\right)\,\,
 \iscol{1}{2} \,\,\,{\bf Sp}(p_1,p_2) \,
{\cal M}_{n-1} \left( P, \{p_j\}\right)\, ,
\end{align}
where one of the partons in ${\cal M}_{n-1} \left( P, \{p_j\}\right)$ replaces the collinear pair, and has a colour charge 
${\bf T}={\bf T}_1+{\bf T}_2$ and momentum \hbox{$P=p_1+p_2$}, while the remaining $(n-2)$ partons $\{p_j\}$ are the non-collinear ones in the original amplitude, which we refer to as ``the rest of the process'' below. The splitting amplitude ${\bf Sp}\left( p_1, p_2 \right)$ is an operator in colour space which captures the singular terms for $P^2\to 0$. 
All elements in Eq.~\eqref{Sp_M} have infrared singularities, and these must clearly be related. Furthermore, ${\bf Sp}$ is expected to only depend on the quantum numbers of the collinear pair~\cite{SplittingAmplitudeRefs} to all orders in perturbation theory. Hence also its soft anomalous dimension,
\begin{equation}
\Gamma_{\rm \bf Sp}=\left.(\Gamma_n-\Gamma_{n-1})\right\vert_{1\parallel 2} = \Gamma^{\rm dip.}_{\rm {\bf Sp}} +\Delta_{\rm {\bf Sp}}\,,
\end{equation}
must be independent of the momenta and colour degrees of freedom of the rest of the process. 
This property is automatically satisfied for the dipole formula, but it is highly non-trivial for it to persist when quadrupole corrections are present. Indeed, the quadrupole interaction might introduce correlations between the collinear pair and the rest of the process.
In Refs.~\cite{Becher:2009cu,Dixon:2009ur} this property was used to constrain $\Delta_n$, but this was done under the assumption that $C$ in Eq.~\eqref{eq:expected_Delta} vanishes. Given our result for $\Delta_n^{(3)}$, the non-dipole correction to the splitting amplitude at three loops are determined: 
\begin{align}
&\Delta_{\rm \bf Sp}^{(3)}=\left.(\Delta_n^{(3)}-\Delta_{n-1}^{(3)})\right\vert_{1\parallel 2}=-24\, (\zeta_5+2\zeta_2\zeta_3)\,
\left(f^{abe}f^{cde}\Big\{{\rm \bf T}_1^a,  {\rm \bf T}_1^c\Big\}\left\{{\rm \bf T}_2^b,  {\rm \bf T}_2^d\right\} +\frac12C_A^2  {\rm \bf T}_1\cdot {\rm \bf T}_2\right) \,.
\label{eq:Delta_Sp_3}
\end{align}
We note that $\Delta_{\rm \bf Sp}^{(3)}$ only depends on the colour degrees of freedom of the collinear pair, and is entirely independent of the kinematics, and hence fully consistent with general expectations\footnote{We recall that strict collinear factorization is restricted to time-like kinematics with both collinear partons in the final state, but it is violated for space-like splitting~\cite{Catani:2011st}.}~\cite{SplittingAmplitudeRefs}. We emphasise that $\Delta_{\rm \bf Sp}^{(3)}$ is independent of the value of $n$ that was used to compute it. In particular, $\Delta_{\rm \bf Sp}^{(3)}$ agrees with $\Delta_n^{(3)}$ for $n=3$, in agreement with the fact that $\Delta_2^{(3)} = 0$.
Indeed, the fact that the difference in Eq.~\eqref{eq:Delta_Sp_3} is independent of $n$ requires intricate relations between different sets of diagrams and thus provides a highly non-trivial check of the calculation. 

\section{Conclusions}

To conclude, we computed~\cite{Almelid:2015jia,longpaper} all connected graphs contributing to the soft anomalous dimension in multi-parton scattering and determined the first correction going beyond the dipole formula.
We find that such corrections appear at three-loops already for three coloured partons, but they only involve kinematic dependence in amplitudes with at least four coloured partons, when conformally-invariant cross rations can be formed.
The final result is remarkably simple: it is expressed in terms of single-valued harmonic polylogarithms of uniform weight five.  Finally, we recover the expected behaviour of amplitudes in both the Regge limit and in two-particle collinear limits, and make further concrete predictions in both these limits.\\

\end{document}